\documentclass[aps,pra,reprint,superscriptaddress,flushbottom, 
]{revtex4-1}

\usepackage{amsmath}
\usepackage{amssymb}
\usepackage{amsthm}
\usepackage[normalem]{ulem}
\usepackage{cancel}

\usepackage{setspace}
\long\def\symbolfootnote[#1]#2{\begingroup\def\thefootnote{\fnsymbol{footnote}}\footnote[#1]{#2}\endgroup}
\usepackage{amscd}

\usepackage{color}

\usepackage{float}

\newcommand{\bra}[1]{\langle {#1} \vert}
\newcommand{\ket}[1]{\vert {#1} \rangle}

\newcommand{\del}[0]{\partial}

\newcommand{\trans}{^{\sf T}}

\newcommand{\stout}[1]{\ifmmode\cancel{\ensuremath{#1}}\else\sout{#1}\fi}

\newcommand{\etal}{\emph{et al.}~}
\newcommand{\eg}{\emph{e.g.,}~}
\newcommand{\ie}{\emph{i.e.,}~}
\newcommand{\nn}{\nonumber}

\newcommand{\beq}{\begin{align}}
\newcommand{\eeq}{\end{align}}

\usepackage{epsfig}

\usepackage{hyperref}
\usepackage{cleveref} 
\crefformat{equation}{Eq.~(#2#1#3)} 
\crefmultiformat{equation}{Eqs.~(#2#1#3)}{ and~(#2#1#3)}{, (#2#1#3)}{ and~(#2#1#3)}
\Crefformat{equation}{Equation~(#2#1#3)}
\crefformat{section}{Sec.~#2#1#3}
\Crefformat{section}{Section~#2#1#3}
\crefformat{figure}{Fig.~#2#1#3}
\Crefformat{figure}{Figure~#2#1#3}

\begin{document}

\title{Optimality of Gaussian receivers for practical Gaussian distributed sensing}
\author{T.J. Volkoff}
\email{volkoff@konkuk.ac.kr}
\affiliation{Department of Physics, Konkuk University, Seoul 05029, Korea}
\author{Mohan Sarovar}
\email{mnsarov@sandia.gov}
\affiliation{Extreme-scale Data Science and Analytics, Sandia National Laboratories, Livermore, California 94550, USA}

\begin{abstract}
We study the problem of estimating a function of many parameters acquired by sensors that are distributed in space, \eg the spatial gradient of a field. We restrict ourselves to a setting where the distributed sensors are probed with experimentally practical resources, namely, field modes in separable displaced thermal states, and focus on the optimal design of the optical receiver that measures the phase-shifted returning field modes. 
Within this setting, we demonstrate that a locally optimal measurement strategy, \textit{i.e.}, one that achieves the standard quantum limit for all phase­ shift values, is a Gaussian measurement, and moreover, one that is separable. We also demonstrate the utility of adaptive phase measurements for making estimation performance robust in cases where one has little prior information on the unknown parameters. In this setting we identify a regime where it is beneficial to use structured optical receivers that entangle the received modes before measurement. 
\end{abstract}
\maketitle

The technical maturity and low cost of a variety of sensors has made distributed sensor networks ubiquitous \cite{Huang:2008gd}. Such sensor networks are advantageous for extracting and processing a variety of spatially distributed information to achieve tasks such as boundary detection and precise estimation of spatially varying fields. 
With the rapid maturation and miniaturization of a variety of quantum sensing technologies, \eg \cite{Mhaskar:2012gk,Maiwald:2009uv,Aasi:2013jb,Korth:2016gb, Chatzidrosos:2017if,Xin:2018eo,Degen:2017kw}, \emph{distributed quantum sensing} is naturally emerging as a technological possibility. 
However, there are still open questions regarding the extent to which quantum sensors can improve performance for distributed sensing problems. 

In the distributed sensing context, one can have two types of quantum sensors. In the first type, each of the $N$ sensing nodes in a network could operate quantum mechanically, but independently of all other nodes, while in the second type, all sensing nodes could be coherently linked, \eg by sharing an entangled state or by being jointly measured by an entangling measurement. For the first type, any quantum enhancement in performance is the same as in the non-distributed setting since one just has $N$ independent sensors. For the second type, there is potential for a quantum-enhancement for sensing distributed properties due to shared quantum resources, and we will focus on this case here.
In this context, Proctor \etal have recently shown that in a network where the quantum state of each sensing node is dependent on a separate parameter, whether there is a benefit to using quantum resources (such as entanglement across the nodes or an entangling measurement) depends on the form of the distributed quantity one is interested in sensing \cite{Proctor:2018dk}. In particular, they show by computing the quantum Fisher information (QFI), that if the goal is to estimate all parameters, there is no benefit to using quantum resources, but that if the goal is to estimate a global (non-local) function of the parameters, then one can obtain a $1/N$ enhancement in precision by initializing all sensor nodes in a quantum entangled state. Several other recent works have also examined QFI and optimal input states for distributed quantum sensing \cite{Humphreys:2013gz,Yue:2014uz,Gagatsos:2016ff,Knott:2016ig,Ciampini:2016cf,Liu:2016by, Zhang:2017bd, Ge:2017wf}.

While the QFI optimized over input states yields the ultimate bound on asymptotic estimation variance, it can be misleading if the measurements required to achieve this bound are not considered since these measurements may be unfeasible under practical constraints. Moreover, the QFI-optimal input states are usually non-classical (and sometimes entangled) states, and preparing many remote quantum sensors in non-classical states (or probing many sensors with entangled probe states) will be technically challenging in the near-term. 

Motivated by these considerations, in this work we consider a practical variant of the distributed quantum sensing problem, 
and quantify the benefits of using realizable measurements to estimate functions of distributed parameters. In particular, we consider a scenario where $N$ quantum sensors are interrogated by separable, classical states that can be measured jointly after interacting with the sensors, see \cref{fig:schematic}. 
Although such a setting is strictly less powerful than the more general one where one also allows for entangled probe states \cite{Giovannetti:2011jk}, it is more practical in the near-term, where constructing joint measurements is more technically feasible. 
We explicitly construct the optimal Gaussian (including adaptive Gaussian) measurement strategies for 
practical distributed sensing with displaced thermal probe states, and show that separable Gaussian measurements can achieve the standard quantum limit in this setting. 
Finally, we identify a special case where a mismatch in prior information about the distributed parameters yields a benefit to using a structured optical receiver that entangles the received light.

\section{Setting}
\label{sec:setting}
\begin{figure}[t!]
\begin{center}
\includegraphics[scale=0.4]{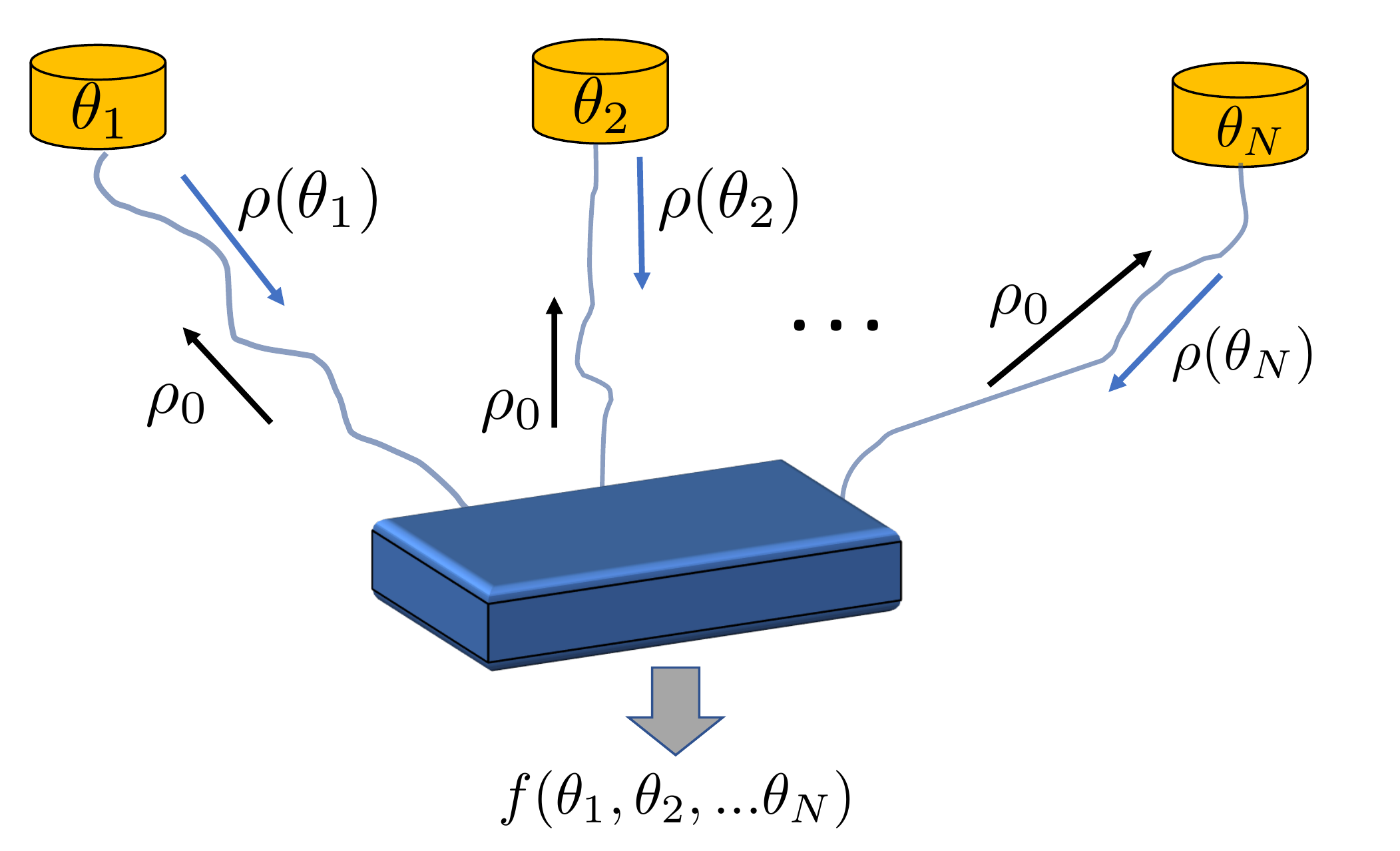}
\caption{\label{fig:schematic} Schematic of the distributed quantum sensing setting considered here. 
Probe states (\eg optical modes) are sent to several distant sensors and return to a central receiver with a parametric dependence ($\theta_i$) on the distributed information to be sensed. We focus on scenarios where some scalar function of all the parameters, $f(\{\theta_i\})$, is the quantity of interest.
}
\end{center}
\end{figure}

 Consider $N$ sensors that are individually probed by $N$ optical probes, each of which is initially in a displaced thermal state and acquires a phase shift $\theta_i$, see \cref{fig:schematic}. The $N$ modes are collected by a receiver, which also has a local phase reference, and the goal is to estimate a function $f(\theta_1, \theta_2, ..., \theta_N)$ of all the parameters. The classical strategy is to measure each mode separately and compute the function $f$ from the measurement results. We ask if performing a joint measurement on the $N$ modes (plus the phase reference mode) is of any benefit. 
Such a setting is relevant to any experimental scenario where information is imprinted in the phase of optical probes. Two examples are: laser phase-shift based range finding \cite{Conde:2017it} and off-resonant optical probing of an array of neutral atoms encoding sensed information in clock state populations \cite{Lodewyck:2009ib}. 

\section{Two-mode, noiseless case}
\label{sec:tmn}
We first consider the case $N=2$ with no propagation loss or measurement noise in order to present the main concepts. The probe state is a two mode displaced thermal state, $\rho_{\rm in}= D(\alpha_{1},\alpha_{2})\rho_{\beta_{1}}\otimes \rho_{\beta_{2}}D^{\dagger}(\alpha_{1},\alpha_{2})$, where 
\begin{align}
D(\alpha_{1},\alpha_{2}):=e^{\sum_{j=1}^{2}\alpha_{j} a_{j}^{\dagger}-\overline{\alpha_{j}}a_{j}},	\nn
\end{align}
is the two-mode displacement operator, 
\begin{align}
\rho_{\beta}:=(1-e^{-\beta})\sum_{n=0}^{\infty}e^{-\beta n}\ket{n}\bra{n},
\end{align}
is a centered, thermal state, and we take $\alpha_{j}\in\mathbb{R}$ for simplicity. The phase shifted state received by the receiver is then $\rho_{\vec{\theta}} = U_{\vec{\theta}}\rho_{\rm in}U_{\vec{\theta}}^{\dagger}$, where 
\begin{align}
	i\ln U_{\vec{\theta}} = \theta_{1}a_{1}^{\dagger}a_{1} + \theta_{2}a_{2}^{\dagger}a_{2}=:H(\vec{\theta}).
\end{align}
Note that $\rho_{\rm in}$ and $\rho_{\vec{\theta}}$ are both two-mode Gaussian states \cite{Weedbrook:2012tz}.

\subsection{The QFI bound}
\label{sec:QFI_bound}
 To motivate the Gaussian measurements considered later, let us first derive the unconstrained optimal question (\ie two element projection-valued measurement) for estimation of the phase difference, $\varphi_{1}:={\theta_{1}-\theta_{2}\over \sqrt{2}}$ between the two modes. We compute the QFI and optimal measurement that saturates it for this case by computing the symmetric logarithmic derivative (SLD). For background on the quantum Cram\'{e}r-Rao bound, the symmetric logarithmic derivative (SLD), and estimation of bosonic Gaussian states, see Ref.~\cite{holevoprob}.
 
 Explicitly calculating derivatives, one finds that
\begin{equation}
\del_{\theta_{j}}\rho_{\vec{\theta}}=ie^{i\theta_{j}}{\alpha_{j}\over N_{j}}a_{j}\rho_{\vec{\theta}} + h.c.,
\end{equation}
where $N_{j}:=\langle a^{\dagger}_{j}a_{j}\rangle_{\rho_{\vec{\theta}}}=(e^{\beta_{j}}-1)^{-1}$. 
However, using the identities
\begin{eqnarray}
a_{j}\rho_{\vec{\theta}}&=&\rho_{\vec{\theta}}\left( e^{-\beta_{j}}(a_{j}-\alpha_{j}e^{-i\theta_{j}})+\alpha_{j}e^{-i\theta_{j}} \right) \nonumber \\ \rho_{\vec{\theta}}a_{j}^{\dagger} &=& \left( e^{-\beta_{j}}(a_{j}^{\dagger}-\alpha_{j}e^{i\theta_{j}})+\alpha_{j}e^{i\theta_{j}} \right)\rho_{\vec{\theta}}
\end{eqnarray}
we can rewrite this derivative as $\del_{\theta_{j}}\rho_{\vec{\theta}}=\rho_{\vec{\theta}}\circ L_{\theta_{j}}$, where
\begin{equation}
L_{\theta_{j}}={\alpha_{j}\over N_{j}+{1\over 2}}\left( ie^{i\theta_{j}}a_{j}+h.c. \right) = L_{\theta_{j}}^{\dagger}
\end{equation}
are the SLD operators in the $\del_{\theta_{j}}$ directions. Here, $\circ$ denotes the Jordan product, \ie $A\circ B \equiv \frac{1}{2}(AB+BA)$. Using the Jacobian to transform the two-dimensional tangent subspace $\text{span}\lbrace L_{\theta_{j}}\rbrace_{j=1,2}$ at $\rho_{\vec{\theta}}$ to the basis $\lbrace L_{\varphi_{j}}\rbrace_{j=1,2}$ gives the SLD with respect to the parameter of concern,
\begin{equation}
L_{\varphi_{1}}={1\over \sqrt{2}}\sum_{j=1}^{2}(-1)^{j+1}{\alpha_{j} \over N_{j}+{1\over 2}}\left( ie^{i\theta_{j}}a_{j} + h.c. \right).
\label{eqn:sldgenclassgauss}
\end{equation}

The QFI is independent of $\theta_{1}$ and $\theta_{2}$ and has the value 
\begin{equation}
	\text{tr}L_{\varphi_{1}}^{2}\rho_{\vec{\theta}}=\sum_{j=1}^{2}{\alpha_{j}^{2}\over N_{j}+{1\over 2}}
\label{eqn:QFI}
\end{equation}
When $N_{1}=N_{2}=0$ this quantity is $2\alpha_{1}^{2}+2\alpha_{2}^{2}$, which is the standard quantum limit (SQL) for estimation of $\varphi_1$ with separable probe states having total intensity $\bar{n}_{\rm tot}=\alpha_{1}^{2}+\alpha_{2}^{2}$ \cite{jarzynawithandwithout}\footnote{As Ref. \cite{jarzynawithandwithout} points out, the SQL for phase difference estimation depends on whether we assume access to a phase reference or not. In our setting, the receiver has a local oscillator that provides a phase reference (\eg has a fixed phase relationship to $\rho_{\rm in}$) and hence the SQL in this setting is $\alpha_1^2 + \alpha_2^2$.}. 
For general two-mode Gaussian states, the SLD for phase difference estimation can also be derived from general formulas for the SLDs on a multimode Gaussian state manifold \cite{adessomulti,serafinibook}. The fact that the SLD for phase difference estimation for probe states of the form $\rho_{\text{in}}$  can be written as a linear function of canonical boson operators $a_{j}$ and $a^{\dagger}_{j}$ is a consequence of the fact that the covariance matrix of $\rho_{\text{in}}$ is invariant under local rotations, i.e., under the adjoint action of $O(2)\times O(2)$.

Furthermore, for $N_{1}=N_{2}=0$ and $\alpha_{1}=\alpha_{2}=:\alpha$, the SLD $L_{\varphi_{1}}$ can be replaced by a rank 2 self-adjoint operator $PL_{\varphi_{1}}P$ given by projecting $L_{\varphi_{1}}$ on both sides, such that 
\begin{align}
PL_{\varphi_{1}}P &= 2\del_{\varphi_{1}}U_{\vec{\theta}} ( \ket{\alpha}\otimes \ket{\alpha}\bra{\alpha}\otimes \bra{\alpha})U_{\vec{\theta}}^{\dagger} \nn \\
&= i\Big[a^{*}_{1}a_{1}-a^{*}_{2}a_{2}, \nn \\
&~~~~~~~~~~ \ket{\alpha e^{-i\theta_{1}}}\otimes \ket{\alpha e^{-i\theta_{2}}}\bra{\alpha e^{-i\theta_{1}}}\otimes \bra{\alpha e^{-i\theta_{2}}}\Big]. \nn	
\end{align}
 See Appendix \ref{sec:app1} for generic construction of $P$. Then at $\theta_{1}=\theta_{2}=0$, the optimal question is given by the spectral projections $\lbrace \ket{\xi_{\pm}}\bra{\xi_{\pm}} \rbrace$ of $PL_{\varphi_{1}}P\vert_{ \vec{\theta}=0 }$, where $\ket{\xi_{\pm}} := {\ket{e_{1}} \pm i\ket{e_{2}} \over \sqrt{2}}$ and
\begin{eqnarray}
\ket{e_{1}} := \ket{\alpha} \otimes \ket{\alpha},
~~~~\ket{e_{2}}:=  {{ \left( a_{2}^{\dagger}a_{2}-a_{1}^{\dagger}a_{1} \right) \ket{\alpha} \otimes \ket{\alpha} } \over \alpha  \sqrt{2} } \nn
\end{eqnarray}
are orthogonal states. The state $\ket{e_{2}}$ is a superposition of photon-added coherent states. Therefore, implementation of the optimal question for estimation of $\varphi_{1}$ requires projection onto entangled non-Gaussian states, suggesting that highly non-trivial quantum resources are necessary to achieve the SQL. However, we proceed to show in Section \ref{sec:sepsec} that a separable Gaussian measurement can approach the same performance. 
For classical, pure Gaussian probe states, \emph{i.e.}, coherent states, this is a consequence of the fact that for pure states, implementation of the projective measurement defined by the SLD is sufficient, but not necessary to saturate the quantum Cram\'{e}r-Rao bound \cite{braunsteincaves}.

\subsection{Restricting to Gaussian measurements}
We denote by $z:=(x_{1},y_{1},x_{2},y_{2})\trans$ the column vector of coordinates on $\mathbb{R}^{4}$, $R:=(q_{1},p_{1},q_{2},p_{2})$ the row vector of canonical observables that satisfy the Heisenberg uncertainty principle $[Rz,Rz']=iz\trans\Delta z' \mathbb{I}_{4}$ for all $z$, $z'\in \mathbb{R}^{4}$ ($\Delta : = \oplus_{j=1}^{2}i\sigma_{y}$ is the standard symplectic form on $\mathbb{R}^{4}$ and we have taken $\hbar =1$), and $W(z):= e^{iRz}$ is a unitary operator that defines the Weyl form of the canonical commutation relations via $W(z)W(z')=e^{-{i\over 2}z\trans\Delta z'}W(z+z')$. $W(z)$ is equal to the two-mode quantum optical displacement operator $D(\alpha_{1},\alpha_{2})$ if one takes 
\begin{align}
	z=(\sqrt{2}\text{Im}\alpha_{1} , -\sqrt{2}\text{Re}\alpha_{1},\sqrt{2}\text{Im}\alpha_{2} , -\sqrt{2}\text{Re}\alpha_{2})\trans. \nn
\end{align}
A Gaussian quantum state $S$ on two modes of the electromagnetic field is associated with a mean vector $m_{S}:=\text{tr}SR$ and a $4\times 4$ covariance matrix $(\Sigma_{S})_{i,j}:=\text{tr}S((R_{i}-m_{i})\circ (R_{j}-m_{j}))$, where $\circ$ denotes the Jordan product.  An energy-constrained Gaussian measurement (ECGM) on two modes is defined by $E\ge 0$ and a positive operator-valued measure 
\begin{align}
	M_{S}(d^{4}z):= W(z)SW(-z) d^{4}z, \nn
\end{align}
with symplectic outcome space $\mathbb{R}^{4}$ such that $S$ is a two-mode, centered Gaussian state (i.e., $m_{S}=\text{tr}SR=(0,0,0,0)$) and $\text{tr}S\sum_{j=1}^{2}a_{j}^{\dagger}a_{j} = E$. Due to the fact that $S$ is centered, the energy constraint can be rewritten ${1\over 2}\text{Tr}\Sigma_{S} - 1 = E$. We note that in order to construct a measurement $\tilde{M}_{S}(d\theta)$ with phase-valued outcomes (\emph{i.e.,} outcomes being measurable subsets of $[0,2\pi)$) which is directly useful for estimation of a relative phase at a certain point in quantum state space, one must push forward the Gaussian measurement $M_{S}(d^{4}z)$ via post-processing of the phase space measurement outcome. However, the Fisher information and optimal measurement depend only on the probe state and the Gaussian state $S$ that defines the ECGM. Note that when $E=0$, this ECGM simply describes a heterodyne measurement, and similary, when $E\rightarrow\infty$, it describes a homodyne measurement. We will refer to these as the heterodyne and homodyne limits, respectively. For $0<E<\infty$, the ECGM prescribes projection onto a state with finite squeezing along some quadrature of a mode, which is practically implemented as an adaptive phase measurement \cite{Wis-1996}. Hence, the parameter $E$ enables us to consider the full class of Gaussian measurements, including adaptive strategies. We stress that we refer to $E$ as an energy constraint in analogy with how this parameter would enter in a description of a Gaussian \emph{state} (in which case, it represents the energy of the state). In the context of Gaussian measurements, this parameter does not represent a physical constraint on energy since $E\rightarrow \infty$ is easily achievable by homodyne measurements.

We consider the single-parameter estimation problem with Cram\'{e}r-Rao bound defined by the Fisher information $\tilde{F}(\rho_{\vec{\theta}})_{1,1}:=(J\trans F(\rho_{\vec{\theta}})J)_{1,1}$, where $F(\rho_{\vec{\theta}})$ is the Fisher information metric on the two-dimensional tangent subspace spanned by $(\del_{\theta_{1}},\del_{\theta_{2}})$ at the probability density $p_{\vec{\theta}}(z):=\text{tr}W(z)SW(-z)\rho_{\vec{\theta}}$, and $J$ is the Jacobian matrix of the transformation from $(\theta_{1},\theta_{2})$ to $(g_{1}(\theta_{1},\theta_{2}),g_{2}(\theta_{1},\theta_{2}))$. The matrix elements of $F(\rho_{\vec{\theta}})$ are defined as:
\begin{eqnarray}
F_{i,j}&:=& \int d^{4}z \, p_{\vec{\theta}}(z)\del_{\theta_{i}}\log p_{\vec{\theta}}(z)\del_{\theta_{j}}\log p_{\vec{\theta}}(z) \nonumber \\ 
&=& \int d^{4}z \, p_{\vec{\theta}}(z)^{-1}\del_{\theta_{i}}p_{\vec{\theta}}(z)\del_{\theta_{j}} p_{\vec{\theta}}(z).
\end{eqnarray}
To calculate $p_{\vec{\theta}}(z)$, we use the expansion of the states over the CCR C$^{*}$-algebra, \emph{e.g.,} $\rho_{\vec{\theta}} = \int {d^{4}z_{1}\over (2\pi)^{2}}\chi_{\rho_{\vec{\theta}}}(z_{1})W(-z_{1})$ where $\chi_{\rho_{\vec{\theta}}}(z_{1}) := e^{-{1\over 2}z_{1}\trans\Sigma_{\rho_{\vec{\theta}}}z_{1} + im_{\rho_{\vec{\theta}}}\trans z_{1}}$ is the characteristic function of $\rho_{\vec{\theta}}$ defined by the covariance matrix $\Sigma_{\rho_{\vec{\theta}}}$ and the mean vector $m_{\rho_{\vec{\theta}}}$. 
This is a multimode generalization of the calculation of single-mode Cram\'{e}r-Rao bound for phase shift estimation with Gaussian measurements reported in Ref.\cite{monrasphase}.
Explicitly,
\begin{widetext}
\begin{eqnarray}
p_{\vec{\theta}}(z) &=& {1\over (2\pi)^{2}}\text{tr} \int {d^{4}z_{1}\over (2\pi)^{2}}  {d^{4}z_{2}\over (2\pi)^{2}}\chi_{\rho_{\vec{\theta}}}(z_{1})\chi_{S}(z_{2})W(-z_{1})W(z)W(-z_{2})W(-z) \nonumber \\ &=&{1\over (2\pi)^{2}}\text{tr} \int {d^{4}z_{1}\over (2\pi)^{2}}  {d^{4}z_{2}\over (2\pi)^{2}}\chi_{\rho_{\vec{\theta}}}(z_{1})\chi_{S}(z_{2})e^{-i\Delta(z_{2},z)}e^{-{i\over 2}\Delta(z_{1},z_{2})}W(-z_{1}-z_{2}) \nonumber \\ &=& {1\over (2\pi)^{2}}\int {d^{4}z_{1}\over (2\pi)^{2}}  \chi_{\rho_{\vec{\theta}}}(z_{1})\chi_{S}(-z_{1})e^{i\Delta(z_{1},z)} \nonumber \\ &=& {1\over (2\pi)^{2}}\left( \text{det}\left( \Sigma_{\rho_{\vec{\theta}}} + \Sigma_{S} \right) \right)^{-1/2} e^{-{1\over 2} \left( m_{\rho_{\vec{\theta}}} - m_{S} - z\trans\Delta \right) \left( \Sigma_{\rho_{\vec{\theta}}} + \Sigma_{S} \right)^{-1}\left( m_{\rho_{\vec{\theta}}} - m_{S} + \Delta z\right)\trans },
\label{eqn:pdensity}
\end{eqnarray}
where, in the third line, we have used $\text{tr}W(z)=(2\pi)^{2}\delta(z)$ for a two mode system. Now, we calculate $\del_{\theta_{1}}p_{\vec{\theta}}(z)$ by using the third line Eq.(\ref{eqn:pdensity}) and a generating function.
\begin{eqnarray}
\del_{\theta_{1}}p_{\vec{\theta}}(z) &=& {1\over (2\pi)^{2}}\int {d^{4}z_{1}\over (2\pi)^{2}}\left[ \vphantom{\sum_{j=1}^{\infty}} \left[ \vphantom{\sum_{j=1}^{\infty}} -{1\over 2} \sum_{m,n=1}^{2}(-i\del_{j_{m}})[\del_{\theta_{1}}\Sigma_{\rho_{\vec{\theta}}}]_{m,n} (-i\del_{j_{n}})  + \sum_{n=1}^{2} i[\del_{\theta_{1}}m_{\rho_{\vec{\theta}}}]_{n}(-i\del_{j_{n}}) \vphantom{\sum_{j=1}^{\infty}} \right] \right. \nonumber \\ && \left. ~~~~~~~~~~~~~~~~~~~~~~ e^{-{1\over 2}z_{1}\trans\Sigma z_{1} + i(m_{\rho_{\vec{\theta}}} - m_{S} - z\trans\Delta) z_{1} }e^{ij\trans z_{1}} \Big\vert_{j=0}\vphantom{\sum_{j=1}^{\infty}} \right] \nonumber \\ &=&
{1\over (2\pi)^{2}\sqrt{ \text{det }\Sigma}}\left[ -{1\over 2}\text{tr}\left( (\del_{\theta_{1}}\Sigma_{\rho_{\vec{\theta}}})\Sigma^{-1} \right) - (\del_{\theta_{1}}m_{\rho_{\vec{\theta}}})\Sigma^{-1}(m_{\rho_{\vec{\theta}}}-m_{S} - z\trans\Delta )\trans \right] p_{\vec{\theta}}(z),
\end{eqnarray}
where the Gaussian integral version of Wick's theorem has been used to get the last line. Now, we perform a final integration over $z$ to get the Fisher metric. Now we explicitly compute the off-diagonal element $F_{1,2}$:
\begin{eqnarray}
F_{1,2}&=& \int d^{4}z \, p_{\vec{\theta}}(z)^{-1}\del_{\theta_{1}}p_{\vec{\theta}}(z)\del_{\theta_{2}}p_{\vec{\theta}}(z) \nonumber \\ &=& {1\over (2\pi)^{2}}\left(\det \Sigma \right)^{-1/2}\nonumber \\ && \int d^{4}z\, \left[ \vphantom{\sum_{j=1}} {1\over 2}\text{tr}\left(  \left(\del_{\theta_{1}}\Sigma_{\rho_{\vec{\theta}}}\right)\Sigma^{-1}\right) + (\del_{\theta_{1}}m_{\rho_{\vec{\theta}}})\Sigma^{-1}( m_{\rho_{\vec{\theta}}}-m_{S} - z\trans\Delta)  \vphantom{\sum_{j=1}} \right] \nonumber \\ && \left[ \vphantom{\sum_{j=1}} {1\over 2}\text{tr}\left(  \left(\del_{\theta_{2}}\Sigma_{\rho_{\vec{\theta}}}\right)\Sigma^{-1}\right) + (\del_{\theta_{2}}m_{\rho_{\vec{\theta}}})\Sigma^{-1}( m_{\rho_{\vec{\theta}}}-m_{S} - z\trans\Delta)  \vphantom{\sum_{j=1}} \right] \nonumber \\ && e^{-{1\over 2} \left( m_{\rho_{\vec{\theta}}} - m_{S} - z\trans\Delta \right) \Sigma ^{-1}\left( m_{\rho_{\vec{\theta}}} - m_{S} + \Delta z\right)\trans }
\end{eqnarray}

Expanding the brackets and noting that: 1) for any $v \in \mathbb{R}^{4}$, and positive $A \in M_{4}(\mathbb{R})$, $\int d^{4}u \, v\trans u\, e^{-{1\over 2}u\trans Au} = 0$, 2) taking $u= m_{\rho_{\vec{\theta}}} - m_{S} + \Delta z$ gives

\begin{eqnarray}
F_{1,2}&=& \left( {1\over 4}\text{tr}\left(  \left(\del_{\theta_{1}}\Sigma_{\rho_{\vec{\theta}}}\right)\Sigma^{-1}\right)\, \text{tr}\left(  \left(\del_{\theta_{2}}\Sigma_{\rho_{\vec{\theta}}}\right)\Sigma^{-1}\right) \right) \nonumber \\ &+& {(\det \Sigma)^{-1/2}\over (2\pi)^{2}}\int d^{4}u \left( (\del_{\theta_{1}}m_{\rho_{\vec{\theta}}})\Sigma^{-1}u\trans \right)\left( (\del_{\theta_{2}}m_{\rho_{\vec{\theta}}})\Sigma^{-1}u\trans \right)e^{-{1\over 2}u\Sigma^{-1}u\trans} . \label{eqn:f12}
\end{eqnarray}

Using the identity
\begin{equation}
{(\det \Sigma)^{-1/2}\over (2\pi)^{2}}\left[ \sum_{n_{1},n_{2}=1}^{2}(L)_{n_{1}}(-i\del_{j_{n_{1}}})(Q)_{n_{2}}(-i\del_{j_{n_{2}}}) \right] \int d^{4}u e^{-{1\over 2}u\trans\Sigma^{-1}u}e^{iu\trans j} \Big\vert_{j=0} = L\Sigma Q\trans
\end{equation}
for row vectors $L$, $Q\in \mathbb{R}^{4}$, allows one to simplify the last line  of Eq.(\ref{eqn:f12}).

\begin{eqnarray}
F_{1,2}&=& \left( {1\over 4}\text{tr}\left(  \left(\del_{\theta_{1}}\Sigma_{\rho_{\vec{\theta}}}\right)\Sigma^{-1}\right)\, \text{tr}\left(  \left(\del_{\theta_{2}}\Sigma_{\rho_{\vec{\theta}}}\right)\Sigma^{-1}\right) \right) +   (\del_{\theta_{1}}m_{\rho_{\vec{\theta}}})\Sigma^{-1}(\del_{\theta_{2}}m_{\rho_{\vec{\theta}}})\trans   \label{eqn:f12final}
\end{eqnarray}

One can similarly compute $F_{1,1}$ and $F_{2,2}$ by just using the appropriate $\del_{\theta_{j}}$.  From the transformation $\tilde{F}_{i,j} = (J\trans FJ)_{i,j}$ that arises from an arbitrary diffeomorphism $\vec{\theta}=(\theta_{1},\theta_{2})\mapsto (g_{1}(\theta_{1},\theta_{2}),g_{2}(\theta_{1},\theta_{2}))$, one finds that
\begin{eqnarray}
\tilde{F}_{i,j}&=& F_{1,1}(\del_{\theta_{i}}g_{1})(\del_{\theta_{j}}g_{1}) +F_{2,2}(\del_{\theta_{i}}g_{2})(\del_{\theta_{j}}g_{2}) + F_{1,2}(\del_{\theta_{i}}g_{1}\del_{\theta_{j}}g_{2} + \del_{\theta_{i}}g_{2}\del_{\theta_{j}}g_{1}).
\label{eqn:expandjacob}
\end{eqnarray} 
Explicitly, $\tilde{F}_{1,1}$ follows immediately from Eq.(\ref{eqn:f12final}) and Eq.(\ref{eqn:expandjacob}), and is given by
\begin{eqnarray}
\tilde{F}_{1,1}(\rho_{\vec{\theta}})&=& {1\over 4}\left( \text{tr}\left( \left(\del_{\theta_{1}}g_{1}\del_{\theta_{1}}\Sigma_{\rho_{\vec{\theta}}} + \del_{\theta_{1}}g_{2}\del_{\theta_{2}}\Sigma_{\rho_{\vec{\theta}}} \right)\Sigma^{-1} \right) \right)^{2}  +   w_{1,1}\Sigma^{-1}w_{1,1}\trans .
\label{eqn:ftildegencoord}
\end{eqnarray} 
where $\Sigma:= \Sigma_{\rho_{\vec{\theta}}}+\Sigma_{S}$, and $w_{1,1}:=(\del_{\theta_{1}}g_{1}\del_{\theta_{1}}m_{\rho_{\vec{\theta}}} + \del_{\theta_{1}}g_{2}\del_{\theta_{2}}m_{\rho_{\vec{\theta}}} ) \in \mathbb{R}^{4}$.
\end{widetext}
We now specialize to the case of phase-difference estimation, $g_{j}={1\over \sqrt{2}}(\theta_{1}+(-1)^{j}\theta_{2})$ for which Eq. (\ref{eqn:ftildegencoord}) becomes $\tilde{F}_{1,1} ={1\over 4}(F_{1,1} + F_{2,2} - 2F_{1,2})$. When the probe state is a two-mode thermal state, \ie of the form $\rho_{\rm in}$, and only when it is so, the covariance matrix $\Sigma_{\rho_{\vec{\theta}}}$ is independent of $\theta_{1}$, $\theta_{2}$. Explicitly, $\Sigma_{\rho_{\vec{\theta}}}=\oplus_{j=1}^{2}(N_{j}+{1\over 2})\mathbb{I}_{2}$, and $m_{\rho_{\vec{\theta}}}=(\sqrt{2}\text{Re}\alpha , \sqrt{2}\text{Im}\alpha,\sqrt{2}\text{Re}\alpha , \sqrt{2}\text{Im}\alpha)V_{\theta_{1}}\oplus V_{\theta_{2}}$, where 
\begin{equation}
	V_{\theta_{j}}=\begin{pmatrix}
  \cos\theta_{j} & \sin\theta_{j}  \\
  -\sin\theta_{j} & \cos\theta_{j}  
 \end{pmatrix}. 
 \label{eqn:vv}
\end{equation}
In this case, $\tilde{F}_{1,1}$ simplifies to
\begin{equation}
\tilde{F}_{1,1} ={1\over 2} ((\del_{\theta_{1}} - \del_{\theta_{2}})m_{\rho_{\vec{\theta}}})\Sigma^{-1}((\del_{\theta_{1}} - \del_{\theta_{2}})m_{\rho_{\vec{\theta}}})\trans.
\label{eqn:ftilde11}
\end{equation}
We specialize to an isothermal ($\beta_{1}=\beta_{2} = \beta$), path-symmetric ($\alpha_{1}=\alpha_{2}=\alpha$) signal, \ie $\rho_{\vec{\theta}}=U_{\vec{\theta}}D(\alpha,\alpha)\rho_{\beta}\otimes \rho_{\beta}D(\alpha,\alpha)^{\dagger}U_{\vec{\theta}}^{\dagger}$, without sacrificing any important features of the problem. See the Appendix for analysis of the $\beta_1\neq \beta_2$ case. We seek to maximize $\tilde{F}_{1,1}$ over $\Sigma_{S}$ in the case that the state $S$ that defines the ECGM is a pure, two-mode Gaussian state, \emph{i.e.,} $\Sigma_{S}={1\over 2}T\trans T$ for $T\in Sp(4,\mathbb{R})$. Under these assumptions, it follows that $\Sigma = (N_{0}+{1\over 2})\mathbb{I}_{4}+{1\over 2}T\trans T$, where $N_{0} :=(e^{\beta}-1)^{-1} $. Because $\Sigma^{-1}>0$, there exists an orthogonal matrix $O$ that takes the eigenvector corresponding to the maximal eigenvalue of $\Sigma^{-1}$ to the direction $(\del_{\theta_{1}} - \del_{\theta_{2}})m_{\rho_{\vec{\theta}}}$. Because $[O,c\mathbb{I}_{4}]=0$ for any constant $c$, where $\mathbb{I}_{4}$ is the unit of $Sp(4,\mathbb{R})$, we may conjugate $\Sigma$ by the adjoint action of $O$ to achieve the maximum value of $\tilde{F}_{1,1}$, \ie
\begin{eqnarray}
&& \max_{\substack{T\in Sp(4,\mathbb{R})\\ {1\over 4}\text{tr}T\trans T -1 = E}} \tilde{F}_{1,1} \nn \\  
&=& \max_{\substack{T\in Sp(4,\mathbb{R})\\ {1\over 4}\text{tr}T\trans T -1 = E}} {1\over 2}  \Vert (\del_{\theta_{1}} - \del_{\theta_{2}})m_{\rho_{\vec{\theta}}}\Vert^{2}\, \Vert \Sigma^{-1}\Vert \nonumber \\ &=& \max_{\substack{T\in Sp(4,\mathbb{R})\\ {1\over 4}\text{tr}T\trans T -1 = E}}  2\alpha^{2}\, \Vert \left( \left( N_{0}+{1\over 2}\right)\mathbb{I} + {1\over 2}T\trans T \right)^{-1}\Vert .
\label{eqn:maxftilde}
\end{eqnarray}
We refer to the quantity in the first line of Eq.(\ref{eqn:maxftilde}), viz., the Fisher information maximized over all Gaussian measurements, as the \emph{Gaussian Fisher information} (GFI), and it is obviously upper bounded by the QFI. 

To calculate the GFI for symmetric, isothermal states of the form $\rho_{\text{in}}$, it follows from the Euler decomposition of $Sp(4,\mathbb{R})$ \cite{safranekfuentes} and the fact that $\Vert O\trans\Sigma^{-1}O\Vert = \Vert \Sigma^{-1} \Vert$ that we may restrict attention to $\Sigma_{S}=\text{diag}(e^{-2r_{1}}/2,e^{2r_{1}}/2,e^{-2r_{2}}/2,e^{2r_{2}}/2)$, $r_{j}\in \mathbb{R}$, such that $\sum_{j=1}^{2}\sinh^{2}r_{j} = E$. We then have that 
\begin{align}
\Vert \Sigma^{-1} \Vert = (N_{0}+{1\over 2}+{1\over 2}e^{-\max \lbrace r_{1},r_{2}\rbrace })^{-1},	
\end{align}
from which it follows that the constrained maximum of $\tilde{F}_{1,1}$ occurs when all the energy is invested into a single mode. The resulting maximum Fisher information is given by 
\begin{equation}
\max_{\substack{T\in Sp(4,\mathbb{R})\\ {1\over 4}\text{tr}T\trans T -1 = E}} \tilde{F}_{1,1} = {2\alpha^{2}\over N_{0}+1+E-\sqrt{E^{2}+E}}
\label{eqn:enerscaling}
\end{equation}
This is the GFI for the phase difference parameter. Note that in the homodyne limit (\ie $E\rightarrow \infty$), this quantity monotonically increases to $4\alpha^2/(2N_0+1)$, which coincides with the QFI, see Eq.(\ref{eqn:QFI}). Hence, the optimal estimation strategy is achievable by a Gaussian measurement.

It remains to identify the ECGM that achieves the optimal value in \cref{eqn:enerscaling}. An arbitrary pure, centered, two mode Gaussian state $S$ can be written as $S=\ket{\Xi}\bra{\Xi}$ with 
\begin{equation}
\ket{\Xi}:=e^{i\sum_{j=1}^{2}\phi_{j}a_{j}^{\dagger}a_{j}}U_{\zeta}e^{\sum_{j=1}^{2}{r_{j}\over 2} (a_{j}^{2} - a_{j}^{\dagger 2})}  \ket{0}\otimes \ket{0},	
\label{eqn:xistate}
\end{equation}
 $r_{j}\in\mathbb{R}$, and $U_{\zeta}:=e^{\zeta a_{1}^{\dagger}a_{2} - \overline{\zeta}a_{2}^{\dagger}a_{1}}$ being a beam-splitter ($\zeta$ is an angle in the closed complex disk with center 0 and radius $\pi/2$) \cite{Dut.Muk.etal-1995}. We set $\phi_{j}=\text{Arg}\zeta=0$ because these parameters do not impact the GFI and hence can be set arbitrarily when defining the optimal measurement. Utilizing this explicit form for the ECGM, the energy constrained maximization of Eq.(\ref{eqn:ftilde11}) at the parameter values $\theta_1=\theta_2=0$ \footnote{While the GFI does not depend on the values of the parameters, the form of the optimal measurement does.} reduces to maximization of  
\begin{equation} 
 2\alpha^{2}\sum_{j=1}^{2} \frac{1+(-1)^{j+1}\sin 2\vert \zeta \vert}{2N_{0}+1+e^{-2r_{j}}}
 \label{eqn:intermed}
\end{equation} 
subject to $\sum_{j=1}^{2}\sinh^{2}(r_{j})=E$. \cref{eqn:intermed} achieves the value in \cref{eqn:enerscaling} when $\vert \zeta \vert = \pi/4$ and when all the energy is invested in squeezing a single mode, \ie $r_{1}=\sinh^{-1}\sqrt{E}$ and $r_{2}=0$. In the homodyne limit, this corresponds to a homodyne measurement of $a_1-a_2$, which is obviously an entangling measurement of the two received modes. In fact, the entanglement entropy in $S=\ket{\Xi}\bra{\Xi}$, takes the value $H(\text{tr}_{2}S) = g({1\over 2}(\sqrt{E+1}-1))$, where $g(x):= (x+1)\log_{2}(x+1)-x\log_{2}x$. The asymptotically optimal Gaussian measurement defined by displacements of $\vert \Xi \rangle \langle \Xi \vert$ coincides with the projection-valued measurement of the observable ${1\over \sqrt{2}}(p_{1}-p_{2})$, which is the measurement defined by the spectral projections of the SLD in Eq.(\ref{eqn:sldgenclassgauss}) for $\alpha_{1}=\alpha_{2}=\alpha \in \mathbb{R}$ and $N_{1}=N_{2}=N_{0}$.

\section{Comparison to separable strategy}
\label{sec:sepsec} 
Having identified the optimal ECGM, we now compare this to the best separable Gaussian strategy, where each received mode is measured separately subject to a total energy constraint. We maximize Eq. (\ref{eqn:ftilde11}) over separable, pure, centered $S$, \ie $S=\ket{\Phi}\bra{\Phi}$  with 
\begin{align}
\ket{\Phi}:=e^{\sum_{j=1}^{2}{r_{j}\over 2} (a_{j}^{2} - a_{j}^{\dagger 2})}  \ket{0}\otimes \ket{0},	
\end{align}
$r_{j}\in\mathbb{R}$, and $\sum_{j=1}^{2}\sinh^{2}r_{j} = E$. 
The resulting quantity is clearly less than or equal to the GFI, and we seek to determine whether it is equal to the GFI.
The state $S$ is a tensor product of single-mode squeezed states and the restriction to real $r_{j}$ is possible because a local rotation of $S$ only decreases the maximum constrained value of $\tilde{F}_{1,1}$. Utilizing this explicit form for the separable ECGM, the maximization of Eq.(\ref{eqn:ftilde11}) reduces to maximization of
$2\alpha^{2}\sum_{j=1}^{2}(2N_{0}+1+e^{-2r_{j}})^{-1}$ subject to $\sum_{j=1}^{2}\sinh^{2}r_{j} = E$. 
While for finite $E$ this quantity is always less than the QFI, $4\alpha^2/(2N_0+1)$, and also less than the optimal value for arbitrary ECGM appearing in Eq. (\ref{eqn:enerscaling}), in the homodyne limit it asymptotes to the QFI. 
Hence, the SQL for estimation of $\varphi_{1}$ with probe states of the form $\rho_{\text{in}}$ is achievable by \emph{separable} homodyne measurements on the two modes. This fact holds even for the case of non-isothermal probe states $\rho_{\text{in}}$ (see Appendix \ref{sec:app2} for a proof). These results emphasize the fact that the optimal Gaussian measurement derived from the SLD in Eq.(\ref{eqn:sldgenclassgauss}) can be post-processed by a rotation, corresponding to an element of the compact subgroup $O(4)$ of $Sp(4,\mathbb{R})$, and still achieve the quantum Cram\'{e}r-Rao bound for estimation of $\varphi_{1}$.
We note that this is particularly striking in the pure, classical Gaussian probe state case where the analysis at the end of section \ref{sec:QFI_bound} based on the projected SLD $PL_{\varphi_{1}}P$, suggested a non-Gaussian, entangling projective measurement to achieve the SQL. This highlights the importance of keeping in mind that
the measurement constructed from SLD eigenstates is sufficient, but not necessary, for
achieving the SQL in the pure state case.

\section{Generalizations} 
In this section we generalize the above calculations to the case of $N$ probe modes and estimation of arbitrary linear functions of the parameters $\theta_i$. But first, we comment on another type of generalization of the above calculations, namely, to include noise. 
The effects of common imperfections in the transmission channel are easily incorporated into the above analysis. Transmission through common media such a fibers and free-space is modeled well by 
compositions of linear bosonic channels that model loss and injection of thermal noise \cite{holevogiov}. These effects simply rescale the amplitude and effective temperature of the received state, $\rho_{\vec{\theta}}$, respectively; \ie $\alpha \rightarrow \eta \alpha$, where $0\leq \eta \leq 1$, and $N_0 \rightarrow N_0 + N_{\rm channel}$.

Now, we consider the generalization to $N$ sensors, each probed by a displaced thermal state that picks up a phase shift $\theta_i$. We define the quantity of interest as the first component of the general linear function $g(\vec{\theta})=(\vec{v}_{1}\cdot \vec{\theta},\ldots ,\vec{v}_{N}\cdot \vec{\theta})$ where $\lbrace \vec{v}_{k} \rbrace_{k=1,\ldots ,N}$ is an orthonormal set in $\mathbb{R}^{N}$.
Instead of working with the full Fisher information matrix $F$ with respect to the tangent space basis $\lbrace {\del \over \del \theta_{j}} \rbrace_{j=1,\ldots ,N}$, we can rotate the system so that the single parameter of interest, i.e., the linear function $g(\vec{\theta})_{1}$, corresponds to the single basis vector $\vec{v}_{1}\cdot \nabla_{\theta}$ for the tangent space.
Specifically, this rotation is given by $\vec{\theta}\mapsto [g(\vec{\theta})_{1},\ldots , g(\vec{\theta})_{N} ]$ and the corresponding Jacobian matrix is  $J:= [\vec{v}_{1},\ldots,\vec{v}_{N}]$. Then, from $\tilde{F} = J\trans FJ$, we get
\begin{equation}
\tilde{F}_{1,1}=(\vec{v}_{1}\cdot \nabla_{\theta})m_{\rho_{\vec{\theta}}} \left( \Sigma_{\rho_{\vec{\theta}}}+\Sigma_{S} \right)^{-1} \left( (\vec{v}_{1}\cdot \nabla_{\theta})m_{\rho_{\vec{\theta}}} \right)\trans 
\label{eqn:f11N}
\end{equation}
which appears in the single-parameter Cram\'{e}r-Rao bound for estimation of $g(\vec{\theta})_{1}$.
We now seek to maximize $\tilde{F}_{1,1}$ subject to the energy constraint $\langle \sum_{j=1}^{N}a_{j}^{\dagger}a_{j} \rangle_{S} = E$; \ie 
\begin{equation}
\max_{\substack{T\in Sp(2N,\mathbb{R}) \\{1\over 4}\text{tr}T\trans T - {N\over 2} = E} } \Vert (\vec{v}_{1}\cdot \nabla_{\theta})m_{\rho_{\vec{\theta}}} \Vert^{2} \Vert ( \Sigma_{\rho_{\vec{\theta}}} + {1\over 2}T\trans T )^{-1}\Vert .
\end{equation}
This can be solved in the same way as the two mode case if we take the isothermal, path symmetric probe state $\rho_{\rm in}= D(\alpha,\ldots , \alpha)\rho_{\beta}^{\otimes N}D(\alpha,\ldots , \alpha)^{\dagger}$, since in this case $\Vert ( \Sigma_{\rho_{\vec{\theta}}} + {1\over 2}T\trans T )^{-1} \Vert = \Vert ( (N_{0} +(1/2))\mathbb{I}_{2N} + \Sigma_{S} )^{-1} \Vert$, and we can assume that $\Sigma_{S}$ is diagonal, \ie $S$ is a tensor product of squeezed states with squeezing of the $q$ or $p$ quadrature only. Clearly, the matrix norm will be maximized if all the squeezing is in one mode (\ie all the energy is used for squeezing), and we have 
\begin{align}
	\Vert ((N_{0} +\frac{1}{2})\mathbb{I}_{2N} + \Sigma_{S} )^{-1} \Vert = (N_{0}+\frac{1}{2} + {e^{-2r}\over 2} )^{-1}, \nn
\end{align}
where $\sinh^{2}r=E$. Rewriting, and using the fact that $m_{\rho_{\vec{\theta}}} = m_{\rho_{\vec{\theta}=0}} \bigoplus_{j=1}^{N} V_{\theta_j}$, where $V_{\theta_{j}}$ is defined in Eq.(\ref{eqn:vv}), gives:

\begin{eqnarray}
\max_{\substack{T\in Sp(2N,\mathbb{R}) \\{1\over 4}\text{tr}T\trans T - {N\over 2} = E} }\tilde{F}_{1,1}&=& {\Vert (\vec{v}_{1}\cdot \nabla_{\theta})m_{\rho_{\vec{\theta}}} \Vert^{2} \over \left( N_{0}+1+E-\sqrt{E^{2}+E} \right)} \nonumber \\ &=& {2\alpha^{2} \over \left( N_{0}+1+E-\sqrt{E^{2}+E} \right) } \label{eqn:isothermprobe} 
\end{eqnarray}
This quantity has no dependence on the number of modes $N$ because of the normalization $\Vert \vec{v}_{1} \Vert=1$. 

To compare this to the GFI when one is limited to separable measurements, we maximize $\tilde{F}_{1,1}$ under the restriction of energy constrained separable measurements. 
For $S$ a separable ECGM, from Eq. (\ref{eqn:f11N}), we have that 
\begin{align}
	\tilde{F}_{1,1}=\sum_{j=1}^{N}2(\vec{v}_{1})_{j}^{2}\alpha^{2}\left( N_{0} + (1/2) +{e^{-2r_{j}}\over 2} \right)^{-1},
\end{align}
where $\sum_{j=1}^{2}\sinh^{2}r_{j}=E$.
When attempting to maximize this, care must be taken in consideration of the vector $\vec{v}_{1}$. In particular, 
the best separable strategy actually depends on the structure of $\vec{v}_1$; if $\vec{v}_1$ is dominated by one entry (the \emph{unbalanced} case), say $(\vec{v}_1)_1$, then its preferable to invest most of the energy available for measurement into measuring the first mode. In contrast, if $\vec{v}_1$ contains entries of almost equal magnitude (the \emph{balanced} case), then the best separable strategy distributes the energy available for measurement among all $N$ modes.

In the unbalanced case,
\begin{align}
&\max_{\substack{S\text{ separable, Gaussian} \\ \langle \sum_{j=1}^{N} a_{j}^{\dagger}a_{j} \rangle_{S} = E}} \tilde{F}_{1,1} = \nn \\
&~~~~~~ 2\alpha^{2} \left( \frac{(\vec{v}_{1})_{1}^{2}}{N_{0} + 1+E-\sqrt{E^{2}+E}} + \sum_{j=2}^{N} \frac{(\vec{v}_{1})_{j}^{2}}{N_{0}+1} \right) \nn
 \label{eqn:isosepmeas}
\end{align}
Note that while this equation does not have an explicit dependence on the number of modes, $N$, there is an implicit dependence on this quantity through $(\vec{v}_1)_1$; namely, since this is assumed to be the largest element of the normalized vector $\vec{v}_1$, its magnitude bounds the number of modes, \ie $N > (1-(\vec{v}_1)_1^2)/(\vec{v}_1)_1^2$.

In the opposite extreme, let us consider $(\vec{v}_{1})_{j}^{2} = 1/N$, $j=1,\ldots ,N$ (the balanced case), which encompasses the case of two-mode phase difference sensing that is considered in previous sections. In this case, the maximum is achieved when the constraint energy is distributed equally for squeezing each mode of the state $S$ that defines the ECGM, and we can show,
\begin{align}
\max_{\substack{S\text{ separable, Gaussian} \\ \langle \sum_{j=1}^{N} a_{j}^{\dagger}a_{j} \rangle_{S} = E}} \tilde{F}_{1,1} =   
\frac{2\alpha^{2}}{ N_{0} + 1+{E\over N}-\sqrt{\left({E\over N}\right)^{2}+{E\over N}}} \nn
\end{align}

We define the ratio of Eq. (\ref{eqn:isothermprobe}) to the maximum achieved by separable strategies, the \emph{entanglement gain} (EG). 
As the homodyne limit is taken ($E\rightarrow \infty$), the entanglement gain asymptotes to $EG^{\rm unbal} \rightarrow  {2N_{0}+2 \over (\vec{v}_{1})_{1}^{2} + 2N_{0}+1 }$ for the unbalanced case, and $EG^{\rm bal} \rightarrow 1$ for the balanced case regardless of $N_{0}$. 
We see that in the general $N$ case also, that separable and entangling Gaussian measurements achieve the same estimation performance in the homodyne limit if the linear function to be estimated has the form $\vec{v}\cdot \vec{\theta}$, where $\vec{v}$ has entries of equal magnitude.

\begin{figure}[t!]
\begin{center}
\includegraphics[scale=0.35]{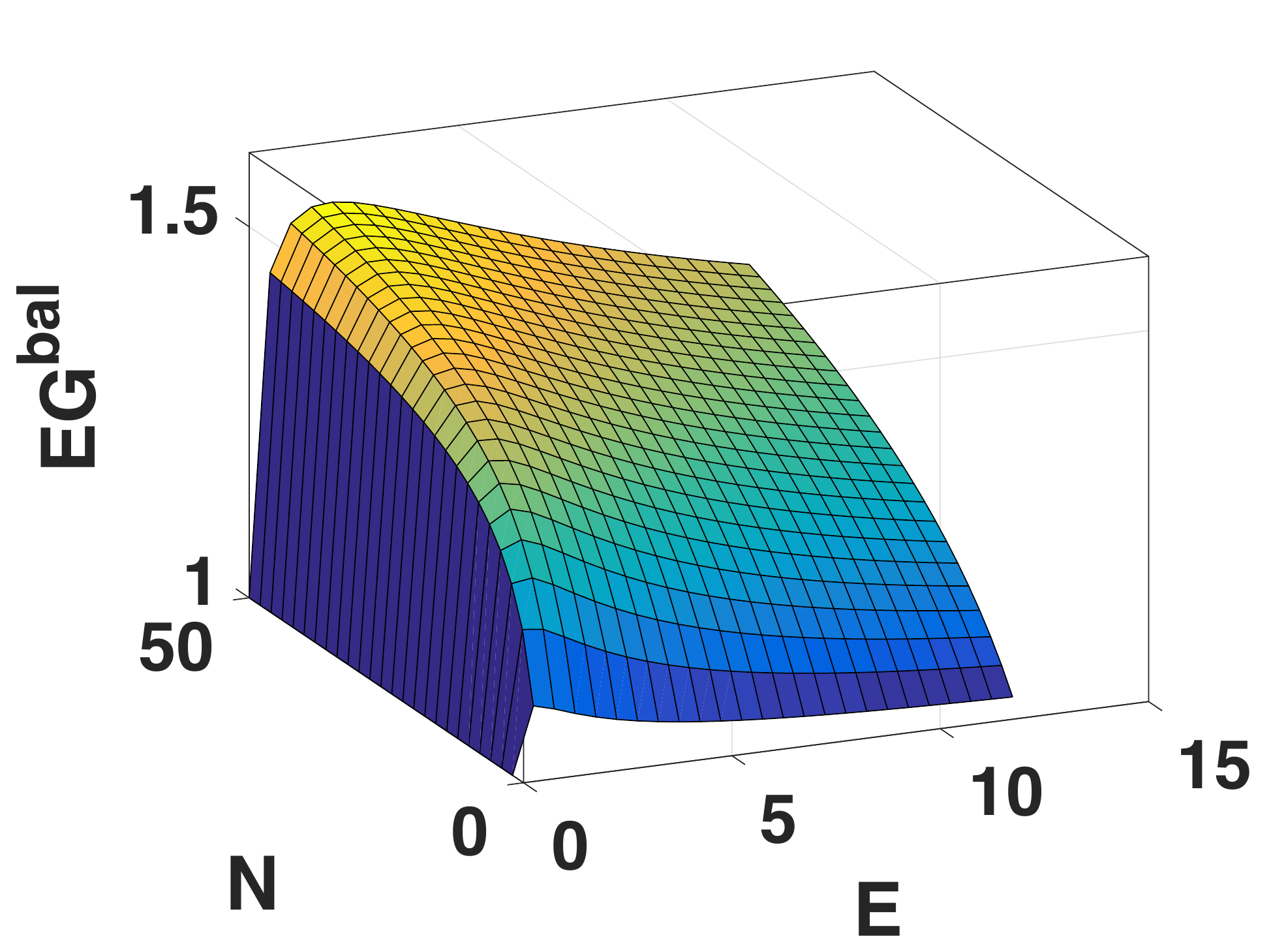}
\quad\quad\quad\quad
\includegraphics[scale=0.35]{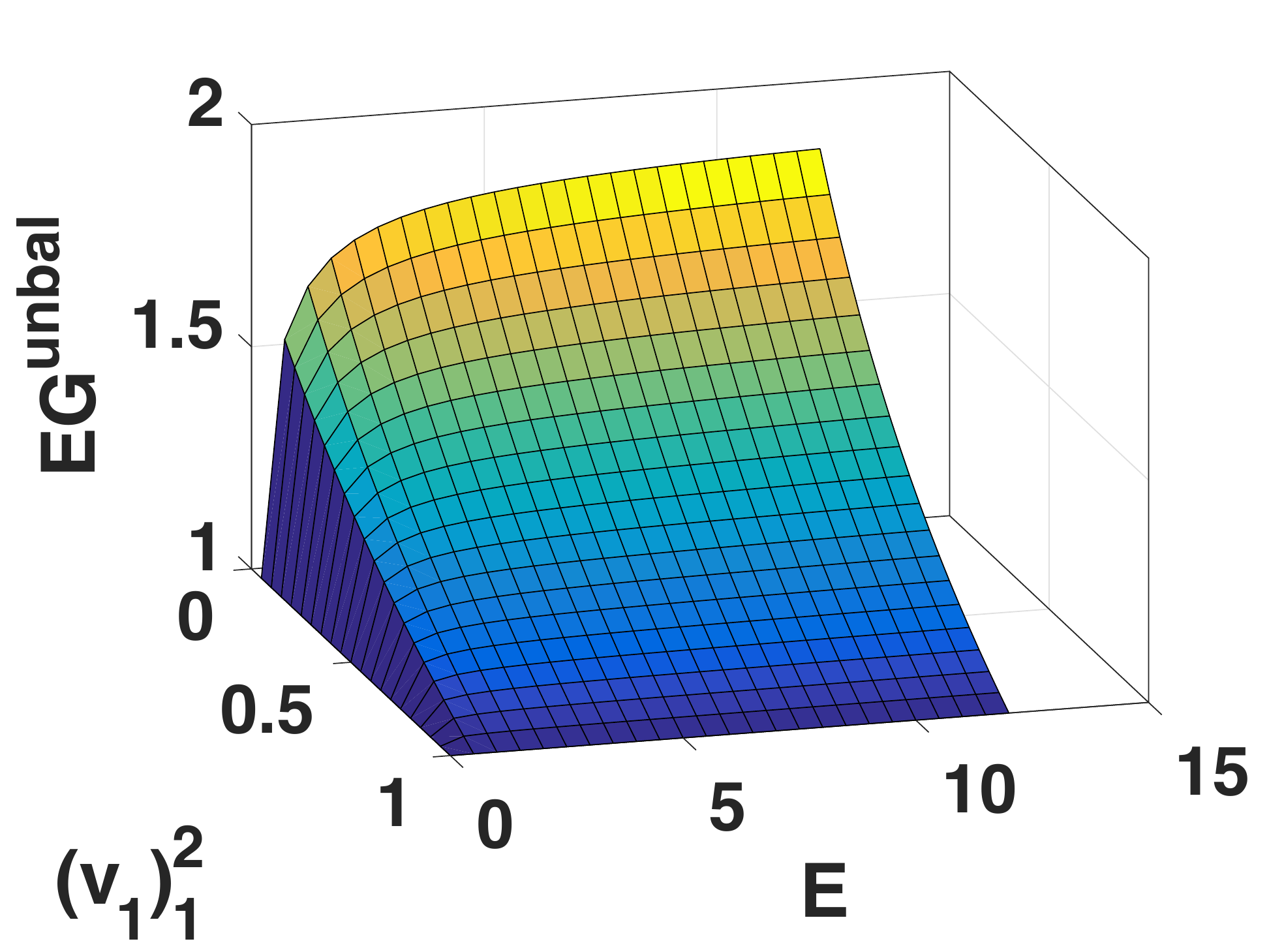}
\caption{\label{fig:EG_unbal} Entanglement gain for the balanced (unbalanced) case as a function of the number of modes, $N$ (the square of the largest element in $\vec{v}_1$, viz., $(\vec{v}_1)_1^2$), and the energy constraint $E$. We set $N_0=0$ in both cases.}
\end{center}
\end{figure}

To appreciate the finite $E$ behavior, in Fig. \ref{fig:EG_unbal} we plot the EG as a function of $E$. For the balanced (unbalanced) case we also show behavior as $N$ ($(\vec{v}_1)_1^2$) is varied. 

\begin{figure}[t]
\begin{center}
\includegraphics[scale=.15]{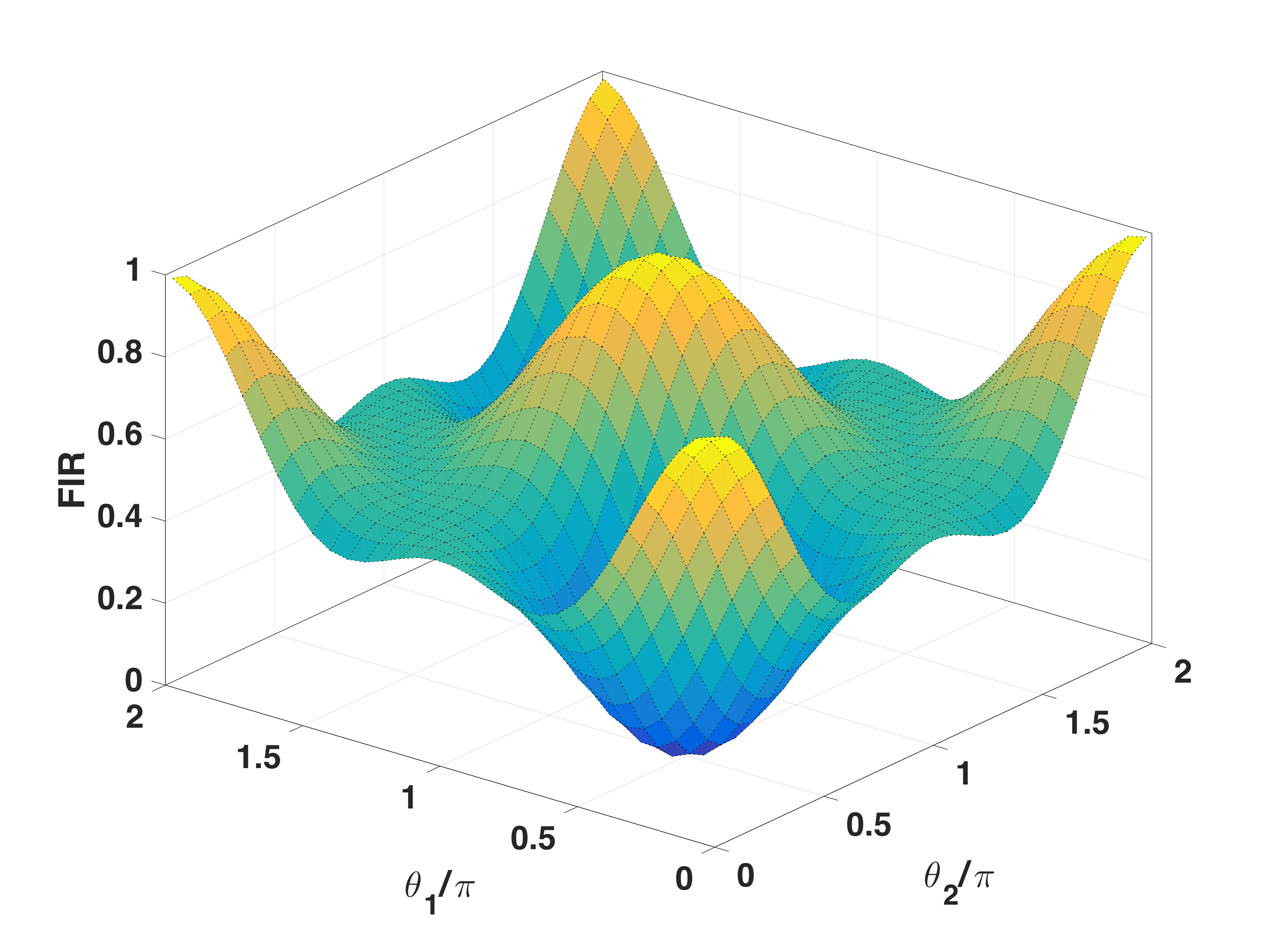}
\caption{\label{fig:fig1}Ratio between the achievable and maximal Fisher information (the FIR) when the measurement used is formulated assuming parameter values $\theta_1=\theta_2=0$, while the true parameter values are indicated on the axes. $\alpha=1, N_0=0$.}
\end{center}
\end{figure} 

\section{Local optimality versus robustness} 
So far we have shown that separable homodyne measurements achieve the optimized Cramer-Rao bound for distributed sensing with displaced thermal state probes; in essence, the best thing to do is the classical strategy of estimating each parameter separately and then computing the function $f(\theta_1,...\theta_N)$. However, it is important to note that the QFI analysis results in locally optimal strategies \cite{DemkowiczDobrzanski:2011ko}. In particular, the form of the optimal measurement is dependent on the values of the parameters $\theta_i$. In the two mode example, the values of $\theta_1,\theta_2$ dictate the local phase parameters $\phi_1,\phi_2$ in the state $\ket{\Xi}$ appearing in Eq.(\ref{eqn:xistate}) that determines the optimal measurement. This is not a practical issue if one has a prior distribution over the parameters that is narrow. However, in cases where this is unavailable, or the prior distribution has broad support (\eg a uniform or maximally uninformative prior) the locally optimal estimation strategy can fail spectacularly. To illustrate this, in \cref{fig:fig1} we plot the ratio between (i) the actual Fisher information achieved when applying the optimal measurement (with $E=10^8$, close to the homodyne limit) formulated for $\theta_1=\theta_2=0$ to a returning state imprinted with different values of $\theta_i$, and (ii) the maximal GFI ($4\alpha^2/(2N_0+1)$). 
If the actual values of the parameters are different from the assumed ones, this Fisher information ratio (FIR) is less that one, and in some cases goes to zero.

We note that a Gaussian strategy that does not suffer from this sensitivity to prior information employs heterodyne measurements for all modes. The Fisher information for heterodyne measurement ($E=0$) is $4\alpha^2/(2N_0+2)$, regardless of whether we allow for entangling, or only separable, measurements. Since this measurement has no dependence on the actual value of the parameters (\ie $\ket{\Xi}=\ket{\Phi} =\ket{0}\otimes \ket{0}$) the Fisher information remains constant regardless of the actual value of the parameters \footnote{It should be noted that although the measurement is independent of the actual phase values, the form of the estimation algorithm (\eg maximal likelihood estimator) that saturates the Cramer-Rao bound may still depend on the phase values.}. However, this lack of sensitivity comes at the cost of a smaller value of Fisher information.  

One way to negotiate this trade-off between estimation precision and robustness is to use adaptive measurements ($0 < E < \infty$) that smoothly interpolate between heterodyne (which prefers no quadrature) and homodyne (which prefers one particular quadrature). In this sense, $E$ can be considered a parameter that quantifies the degree of confidence in the prior information on the parameters. This also suggests a scenario where there is a benefit to using a structured optical receiver. Namely, consider a setting where one is very uncertain about the distribution of the individual parameters $\theta_i$, but has a narrow prior on the collective parameter $f(\theta_1,...\theta_N)$. If one is concerned with minimizing uncertainty in estimation precision (\eg quantified by the variance in Fisher information) then the best separable strategy is to use heterodyne measurements on all modes, in which case the Fisher information is $4\alpha^2/(2N_0+2)$. However, if one employs an entangling measurement that concentrates the collective parameter into a single mode, one can exploit the narrow prior on this parameter and apply a homodyne measurement on this mode to attain the optimized Fisher information for the estimation problem $4\alpha^2/(2N_0+1)$. Although this is only a constant gain in estimation precision it could be beneficial in extremely low-power, low-noise applications ($\alpha^2\ll 1$ and $N_0 \ll 1$).

\section{Conclusions} 
We have analyzed distributed quantum sensing applications where 
displaced, thermal probe fields are imprinted with phase shifts proportional to distributed parameters, and one is interested in estimating a global function of the parameters. 
We proved that a separable, Gaussian measurement is a locally optimal estimation strategy that saturates the SQL for probe states of the form $\rho_{\text{in}}$. Furthermore, we showed that a narrow prior distribution over the parameters is necessary to achieve the optimal precision. Finally, we highlighted a scenario defined by a mismatch between prior information about the individual parameters and a global function of the parameters, where an entangling measurement can yield some benefit, and examined this benefit for a range of Gaussian measurements ($0<E<\infty$). 

We have shown that separable Gaussian estimation is generally the locally optimal strategy for distributed phase estimation with displaced thermal probe states. We expect that this will not be the case for more general probe states, even separable Gaussian states. For example, an interesting problem for future work is to identify the optimal receiver for distributed sensing when separable squeezed states are used as probes.

\begin{acknowledgements}
T.J.V. thanks Yongkyung Kwon for hosting at Konkuk University during the completion of this work. M.S. thanks Grant Biedermann and Tim Proctor for useful discussions about distributed sensing, and Howard Wiseman for discussions about adaptive measurement POVMs. T.J.V. acknowledges support from the National Research Foundation of Korea (NRF) funded by the 
Ministry of Science and ICT (Grant No. 2016H1D3A1908876) and by the Basic Science Research Program through the NRF funded by the Ministry of Education (Grant No. 2015R1D1A1A09056745). Sandia National Laboratories is a multimission laboratory managed and operated by National Technology and Engineering Solutions of Sandia, LLC., a wholly owned subsidiary of Honeywell International, Inc., for the U.S. Department of Energy's National Nuclear Security Administration under contract DE-NA-0003525. This paper describes objective technical results and analysis. Any subjective views or opinions that might be expressed in the paper do not necessarily represent the views of the U.S. Department of Energy or the United States Government.
\end{acknowledgements}
 
\appendix

\section{Construction of $P$}
\label{sec:app1}
For a pure state $\ket{\psi}$, the defining equation of the SLD (in the direction $\del_{\varphi_{1}}$) $\del_{\varphi_{1}}\ket{\psi}\bra{\psi}={1\over 2}\ket{\psi}\bra{\psi}L_{\varphi_{1}} + {1\over 2}L_{\varphi_{1}}\ket{\psi}\bra{\psi}$ combined with the fact that $\langle \psi \vert L_{\varphi_{1}} \vert \psi \rangle=0$ implies that $P(\del_{\varphi_{1}}\ket{\psi}\bra{\psi})P = \del_{\varphi_{1}}\ket{\psi}\bra{\psi}$, where $P$ is the projection to the two-dimensional complex Hilbert space $\overline{\text{span}\lbrace \ket{\psi},L_{\varphi_{1}}\ket{\psi}\rbrace }$ (clearly, $P$ is dependent on $\ket{\psi}$). Then, since $[P,\ket{\psi}\bra{\psi}]=0$, it follows that $\del_{\varphi_{1}}\ket{\psi}\bra{\psi}={1\over 2}\ket{\psi}\bra{\psi}PL_{\varphi_{1}}P + {1\over 2}PL_{\varphi_{1}}P\ket{\psi}\bra{\psi}$. Calculation of the spectral projections of $PL_{\varphi_{1}}P$ amounts to diagonalization of a $2\times 2$ matrix.

\begin{figure}[hbt]
\includegraphics[scale=.15]{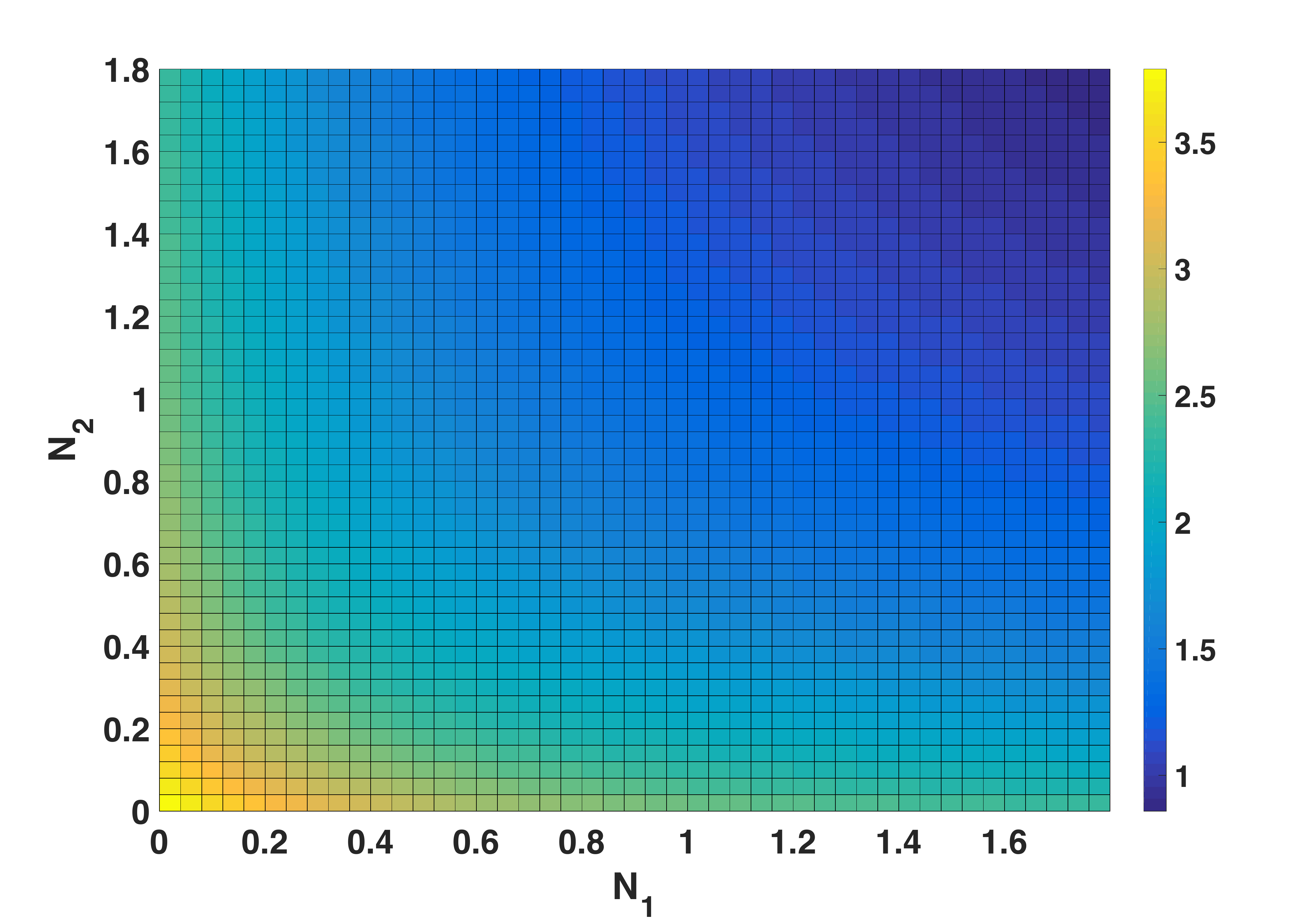}
\caption{\label{fig:sm_noniso} Maximal value of $\tilde{F}_{1,1}$ for probe state $\rho_{\rm in}= D(\alpha,\alpha)\rho_{\beta_{1}}\otimes \rho_{\beta_{2}}D^{\dagger}(\alpha,\alpha)$, with $\alpha =1$. The average thermal photon numbers $N_i = (e^{\beta_i}-1)^{-1}$ are swept across the two axes. The ECGM is optimized subject to the energy constraint $E=4$. $\tilde{F}_{1,1}$ is symmetric about $N_{1}=N_{2}$.}
\end{figure} 

\section{Non-isothermal probe states}
\label{sec:app2}

In the main text, we focus on the case of probe states $\rho_{\text{in}}$ that satisfy the isothermality condition (\ie the inverse temperature of all modes is $\beta = \ln {N_{0}+1\over N_{0}}$). In this section we compute the GFI for non-isothermal states for completeness and then specialize to the case of separable Gaussian measurements and show that for non-isothermal states, such measurements are sufficient to saturate the SQL given by Eq.(\ref{eqn:QFI}).

Consider the $N=2$ case, and path-symmetric, non-isothermal ($\alpha_1=\alpha_2=\alpha$ and $\beta_{1}\neq \beta_{2}$) probes for phase difference estimation. Because $\Sigma_{\rho_{\vec{\theta}}}$ is no longer a constant multiple of the identity matrix, the arguments leading to Eq.(\ref{eqn:maxftilde}) in the main text cannot be applied. In this case, it is most convenient to carry out constrained numerical optimization of Eq.(\ref{eqn:ftilde11}) over states $S$ defined by $S=\ket{\Xi}\bra{\Xi}$ with $\ket{\Xi}$ as defined in the main text, and Fig. \ref{fig:sm_noniso} presents the results of this calculation.  It is clear from this data that the maximal value of $\tilde{F}_{1,1}$ decreases most rapidly for uniform probe state noise. The entanglement entropy of the optimal $S$ (not shown) satisfies the following properties: 1) it is constant along the $N_{1}=N_{2}$ line and in agreement with the value $g({1\over 2}(\sqrt{E +1} -1))$, $g(x):=(x+1)\log_{2}(x+1)-x\log_{2}x$, for all values of $E$ as discussed in the main text and 2) it monotonically decreases from the  $N_{1}=N_{2}$ value along the quarter circle of radius $\sqrt{N_{1}^{2}+N_{2}^{2}}$.

We now proceed to demonstrate that a separable Gaussian measurement corresponding to a local homodyne measurement is sufficient to achieve the standard quantum limit in Eq.(\ref{eqn:QFI}) for the probe states $\rho_{in}$ defined in Section \ref{sec:tmn}. First, consider estimation of $\varphi_{1}$ at the point $\theta_{1}=\theta_{2}=0$, so that the value of the estimand is $\varphi_{1}=0$. One finds that $(\del_{\theta_{1}}-\del_{\theta_{2}})m_{\rho_{\vec{\theta}}}\vert_{\vec{\theta}=0}= (0,\sqrt{2}\alpha_{1},0,-\sqrt{2}\alpha_{2} )$, where $\alpha_{1}$, $\alpha_{2}\in \mathbb{R}$. Taking, as in Section \ref{sec:sepsec}, $S=\ket{\Phi}\bra{\Phi}$, direct calculation of $\tilde{F}_{1,1}$ in Eq.(\ref{eqn:ftilde11}) clearly shows that there are $r_{1}$, $r_{2}\in\mathbb{R}$ such that Eq.(\ref{eqn:QFI}) is obtained for $E\rightarrow \infty$. All that remains is to show that Eq.(\ref{eqn:QFI}) is achievable for all parameter values $\vec{\theta}$. To proceed, note that for any $\vec{\theta}$,
\begin{equation}
(\del_{\theta_{1}}-\del_{\theta_{2}})m_{\rho_{\vec{\theta}}} = (\del_{\theta_{1}}-\del_{\theta_{2}})m_{\rho_{\vec{\theta}}} \vert_{\vec{\theta}=0}\left( V_{\theta_{1}}\oplus V_{\theta_{2}} \right),
\label{eqn:bb}
\end{equation}
where $V_{\theta_{j}}$ is defined in Eq.(\ref{eqn:vv}). Eq.(\ref{eqn:bb}) implies that $ (\del_{\theta_{1}}-\del_{\theta_{2}})m_{\rho_{\vec{\theta}}}$ is on an $O(2)\times O(2)$ orbit that passes through $ (\del_{\theta_{1}}-\del_{\theta_{2}})m_{\rho_{\vec{\theta}}} \vert_{\vec{\theta}=0}$.
Since $[\Sigma_{\rho_{\vec{\theta}}},A\oplus B ] =0$ for any $A$, $B\in O(2)$, it follows that if one takes the separable pure state $S':= U_{\vec{\theta}}^{\dagger} \ket{\Phi}\bra{\Phi} U_{\vec{\theta}}$ to define the Gaussian measurement, with $\ket{\Phi}$ defined with the optimal $r_{1}$, $r_{2}$ values for the $\vec{\theta}=0$ case, then
\begin{widetext}
\begin{eqnarray}
\tilde{F}_{1,1}\vert_{\vec{\theta}=(\theta_{1},\theta_{2})}&=& {1\over 2} ((\del_{\theta_{1}} - \del_{\theta_{2}})m_{\rho_{\vec{\theta}}})\Sigma^{-1}((\del_{\theta_{1}} - \del_{\theta_{2}})m_{\rho_{\vec{\theta}}})\trans \nonumber \\
&=&{1\over 2} ((\del_{\theta_{1}} - \del_{\theta_{2}})m_{\rho_{\vec{\theta}}})\left( \Sigma_{\rho_{\vec{\theta}}} + \Sigma_{S'} \right)^{-1} ((\del_{\theta_{1}} - \del_{\theta_{2}})m_{\rho_{\vec{\theta}}})\trans \nonumber \\
&=& {1\over 2} ((\del_{\theta_{1}} - \del_{\theta_{2}})m_{\rho_{\vec{\theta}}}) \vert_{\vec{\theta}=0} \left( V_{\theta_{1}}\oplus V_{\theta_{2}} \right)\cdot \nonumber \\ &{}& \left( \Sigma_{\rho_{\vec{\theta}}} + {1\over 2}\left( V_{\theta_{1}}^{-1}\oplus V_{\theta_{2}}^{-1} \right) T\trans T \left( V_{\theta_{1}}\oplus V_{\theta_{2}} \right) \right) \nonumber \\ &{}& \left( (\del_{\theta_{1}} - \del_{\theta_{2}})m_{\rho_{\vec{\theta}}}\vert_{\vec{\theta}=0}  \left( V_{\theta_{1}}\oplus V_{\theta_{2}} \right) \right) \trans \nonumber \\
&=& \tilde{F}_{1,1}\vert_{\vec{\theta}=0}
\end{eqnarray}
\end{widetext}
where, in the second line, ${1\over 2}T\trans T$ is the covariance matrix of $\ket{\Phi}$. We have already shown that Eq.(\ref{eqn:QFI}) is attained at $\vec{\theta}=0$, so the proof is finished.

\bibliography{phasebib.bib}

\begin{thebibliography}{35}%
\makeatletter
\providecommand \@ifxundefined [1]{%
 \@ifx{#1\undefined}
}%
\providecommand \@ifnum [1]{%
 \ifnum #1\expandafter \@firstoftwo
 \else \expandafter \@secondoftwo
 \fi
}%
\providecommand \@ifx [1]{%
 \ifx #1\expandafter \@firstoftwo
 \else \expandafter \@secondoftwo
 \fi
}%
\providecommand \natexlab [1]{#1}%
\providecommand \enquote  [1]{``#1''}%
\providecommand \bibnamefont  [1]{#1}%
\providecommand \bibfnamefont [1]{#1}%
\providecommand \citenamefont [1]{#1}%
\providecommand \href@noop [0]{\@secondoftwo}%
\providecommand \href [0]{\begingroup \@sanitize@url \@href}%
\providecommand \@href[1]{\@@startlink{#1}\@@href}%
\providecommand \@@href[1]{\endgroup#1\@@endlink}%
\providecommand \@sanitize@url [0]{\catcode `\\12\catcode `\$12\catcode
  `\&12\catcode `\#12\catcode `\^12\catcode `\_12\catcode `\%12\relax}%
\providecommand \@@startlink[1]{}%
\providecommand \@@endlink[0]{}%
\providecommand \url  [0]{\begingroup\@sanitize@url \@url }%
\providecommand \@url [1]{\endgroup\@href {#1}{\urlprefix }}%
\providecommand \urlprefix  [0]{URL }%
\providecommand \Eprint [0]{\href }%
\providecommand \doibase [0]{http://dx.doi.org/}%
\providecommand \selectlanguage [0]{\@gobble}%
\providecommand \bibinfo  [0]{\@secondoftwo}%
\providecommand \bibfield  [0]{\@secondoftwo}%
\providecommand \translation [1]{[#1]}%
\providecommand \BibitemOpen [0]{}%
\providecommand \bibitemStop [0]{}%
\providecommand \bibitemNoStop [0]{.\EOS\space}%
\providecommand \EOS [0]{\spacefactor3000\relax}%
\providecommand \BibitemShut  [1]{\csname bibitem#1\endcsname}%
\let\auto@bib@innerbib\@empty
\bibitem [{\citenamefont {Huang}\ \emph {et~al.}(2008)\citenamefont {Huang},
  \citenamefont {Hsieh},\ and\ \citenamefont {Sandnes}}]{Huang:2008gd}%
  \BibitemOpen
  \bibfield  {author} {\bibinfo {author} {\bibfnamefont {Y.-M.}\ \bibnamefont
  {Huang}}, \bibinfo {author} {\bibfnamefont {M.-Y.}\ \bibnamefont {Hsieh}}, \
  and\ \bibinfo {author} {\bibfnamefont {F.~E.}\ \bibnamefont {Sandnes}},\ }in\
  \href@noop {} {\emph {\bibinfo {booktitle} {Sensors}}}\ (\bibinfo
  {publisher} {Springer, Berlin},\ \bibinfo {year} {2008})\ pp.\ \bibinfo
  {pages} {199--219}\BibitemShut {NoStop}%
\bibitem [{\citenamefont {Mhaskar}\ \emph {et~al.}(2012)\citenamefont
  {Mhaskar}, \citenamefont {Knappe},\ and\ \citenamefont
  {Kitching}}]{Mhaskar:2012gk}%
  \BibitemOpen
  \bibfield  {author} {\bibinfo {author} {\bibfnamefont {R.}~\bibnamefont
  {Mhaskar}}, \bibinfo {author} {\bibfnamefont {S.}~\bibnamefont {Knappe}}, \
  and\ \bibinfo {author} {\bibfnamefont {J.}~\bibnamefont {Kitching}},\
  }\href@noop {} {\bibfield  {journal} {\bibinfo  {journal} {Applied Physics
  Letters}\ }\textbf {\bibinfo {volume} {101}},\ \bibinfo {pages} {241105}
  (\bibinfo {year} {2012})}\BibitemShut {NoStop}%
\bibitem [{\citenamefont {Maiwald}\ \emph {et~al.}(2009)\citenamefont
  {Maiwald}, \citenamefont {Leibfried}, \citenamefont {Britton}, \citenamefont
  {Bergquist}, \citenamefont {Leuchs},\ and\ \citenamefont
  {Wineland}}]{Maiwald:2009uv}%
  \BibitemOpen
  \bibfield  {author} {\bibinfo {author} {\bibfnamefont {R.}~\bibnamefont
  {Maiwald}}, \bibinfo {author} {\bibfnamefont {D.}~\bibnamefont {Leibfried}},
  \bibinfo {author} {\bibfnamefont {J.}~\bibnamefont {Britton}}, \bibinfo
  {author} {\bibfnamefont {J.~C.}\ \bibnamefont {Bergquist}}, \bibinfo {author}
  {\bibfnamefont {G.}~\bibnamefont {Leuchs}}, \ and\ \bibinfo {author}
  {\bibfnamefont {D.~J.}\ \bibnamefont {Wineland}},\ }\href@noop {} {\bibfield
  {journal} {\bibinfo  {journal} {Nat. Phys.}\ }\textbf {\bibinfo {volume}
  {5}},\ \bibinfo {pages} {551} (\bibinfo {year} {2009})}\BibitemShut {NoStop}%
\bibitem [{\citenamefont {Aasi}\ \emph {et~al.}(2013)\citenamefont {Aasi} \emph
  {et~al.}}]{Aasi:2013jb}%
  \BibitemOpen
  \bibfield  {author} {\bibinfo {author} {\bibfnamefont {J.}~\bibnamefont
  {Aasi}} \emph {et~al.},\ }\href@noop {} {\bibfield  {journal} {\bibinfo
  {journal} {Nature Photonics}\ }\textbf {\bibinfo {volume} {7}},\ \bibinfo
  {pages} {613} (\bibinfo {year} {2013})}\BibitemShut {NoStop}%
\bibitem [{\citenamefont {Korth}\ \emph {et~al.}(2016)\citenamefont {Korth},
  \citenamefont {Strohbehn}, \citenamefont {Tejada}, \citenamefont {Andreou},
  \citenamefont {Kitching}, \citenamefont {Knappe}, \citenamefont {Lehtonen},
  \citenamefont {London},\ and\ \citenamefont {Kafel}}]{Korth:2016gb}%
  \BibitemOpen
  \bibfield  {author} {\bibinfo {author} {\bibfnamefont {H.}~\bibnamefont
  {Korth}}, \bibinfo {author} {\bibfnamefont {K.}~\bibnamefont {Strohbehn}},
  \bibinfo {author} {\bibfnamefont {F.}~\bibnamefont {Tejada}}, \bibinfo
  {author} {\bibfnamefont {A.~G.}\ \bibnamefont {Andreou}}, \bibinfo {author}
  {\bibfnamefont {J.}~\bibnamefont {Kitching}}, \bibinfo {author}
  {\bibfnamefont {S.}~\bibnamefont {Knappe}}, \bibinfo {author} {\bibfnamefont
  {S.~J.}\ \bibnamefont {Lehtonen}}, \bibinfo {author} {\bibfnamefont {S.~M.}\
  \bibnamefont {London}}, \ and\ \bibinfo {author} {\bibfnamefont
  {M.}~\bibnamefont {Kafel}},\ }\href@noop {} {\bibfield  {journal} {\bibinfo
  {journal} {Journal of Geophysical Research: Space Physics}\ }\textbf
  {\bibinfo {volume} {121}},\ \bibinfo {pages} {7870} (\bibinfo {year}
  {2016})}\BibitemShut {NoStop}%
\bibitem [{\citenamefont {Chatzidrosos}\ \emph {et~al.}(2017)\citenamefont
  {Chatzidrosos}, \citenamefont {Wickenbrock}, \citenamefont {Bougas},
  \citenamefont {Leefer}, \citenamefont {Wu}, \citenamefont {Jensen},
  \citenamefont {Dumeige},\ and\ \citenamefont {Budker}}]{Chatzidrosos:2017if}%
  \BibitemOpen
  \bibfield  {author} {\bibinfo {author} {\bibfnamefont {G.}~\bibnamefont
  {Chatzidrosos}}, \bibinfo {author} {\bibfnamefont {A.}~\bibnamefont
  {Wickenbrock}}, \bibinfo {author} {\bibfnamefont {L.}~\bibnamefont {Bougas}},
  \bibinfo {author} {\bibfnamefont {N.}~\bibnamefont {Leefer}}, \bibinfo
  {author} {\bibfnamefont {T.}~\bibnamefont {Wu}}, \bibinfo {author}
  {\bibfnamefont {K.}~\bibnamefont {Jensen}}, \bibinfo {author} {\bibfnamefont
  {Y.}~\bibnamefont {Dumeige}}, \ and\ \bibinfo {author} {\bibfnamefont
  {D.}~\bibnamefont {Budker}},\ }\href@noop {} {\bibfield  {journal} {\bibinfo
  {journal} {Physical Review Applied}\ }\textbf {\bibinfo {volume} {8}},\
  \bibinfo {pages} {044019} (\bibinfo {year} {2017})}\BibitemShut {NoStop}%
\bibitem [{\citenamefont {Xin}\ \emph {et~al.}(2018)\citenamefont {Xin},
  \citenamefont {Leong}, \citenamefont {Chen},\ and\ \citenamefont
  {Lan}}]{Xin:2018eo}%
  \BibitemOpen
  \bibfield  {author} {\bibinfo {author} {\bibfnamefont {M.}~\bibnamefont
  {Xin}}, \bibinfo {author} {\bibfnamefont {W.~S.}\ \bibnamefont {Leong}},
  \bibinfo {author} {\bibfnamefont {Z.}~\bibnamefont {Chen}}, \ and\ \bibinfo
  {author} {\bibfnamefont {S.-Y.}\ \bibnamefont {Lan}},\ }\href@noop {}
  {\bibfield  {journal} {\bibinfo  {journal} {Science Advances}\ }\textbf
  {\bibinfo {volume} {4}},\ \bibinfo {pages} {e1701723} (\bibinfo {year}
  {2018})}\BibitemShut {NoStop}%
\bibitem [{\citenamefont {Degen}\ \emph {et~al.}(2017)\citenamefont {Degen},
  \citenamefont {Reinhard},\ and\ \citenamefont {Cappellaro}}]{Degen:2017kw}%
  \BibitemOpen
  \bibfield  {author} {\bibinfo {author} {\bibfnamefont {C.~L.}\ \bibnamefont
  {Degen}}, \bibinfo {author} {\bibfnamefont {F.}~\bibnamefont {Reinhard}}, \
  and\ \bibinfo {author} {\bibfnamefont {P.}~\bibnamefont {Cappellaro}},\
  }\href@noop {} {\bibfield  {journal} {\bibinfo  {journal} {Rev. Mod. Phys.}\
  }\textbf {\bibinfo {volume} {89}},\ \bibinfo {pages} {035002} (\bibinfo
  {year} {2017})}\BibitemShut {NoStop}%
\bibitem [{\citenamefont {Proctor}\ \emph {et~al.}(2018)\citenamefont
  {Proctor}, \citenamefont {Knott},\ and\ \citenamefont
  {Dunningham}}]{Proctor:2018dk}%
  \BibitemOpen
  \bibfield  {author} {\bibinfo {author} {\bibfnamefont {T.~J.}\ \bibnamefont
  {Proctor}}, \bibinfo {author} {\bibfnamefont {P.~A.}\ \bibnamefont {Knott}},
  \ and\ \bibinfo {author} {\bibfnamefont {J.~A.}\ \bibnamefont {Dunningham}},\
  }\href@noop {} {\bibfield  {journal} {\bibinfo  {journal} {Phys. Rev. Lett.}\
  }\textbf {\bibinfo {volume} {120}},\ \bibinfo {pages} {080501} (\bibinfo
  {year} {2018})}\BibitemShut {NoStop}%
\bibitem [{\citenamefont {Humphreys}\ \emph {et~al.}(2013)\citenamefont
  {Humphreys}, \citenamefont {Barbieri}, \citenamefont {Datta},\ and\
  \citenamefont {Walmsley}}]{Humphreys:2013gz}%
  \BibitemOpen
  \bibfield  {author} {\bibinfo {author} {\bibfnamefont {P.~C.}\ \bibnamefont
  {Humphreys}}, \bibinfo {author} {\bibfnamefont {M.}~\bibnamefont {Barbieri}},
  \bibinfo {author} {\bibfnamefont {A.}~\bibnamefont {Datta}}, \ and\ \bibinfo
  {author} {\bibfnamefont {I.~A.}\ \bibnamefont {Walmsley}},\ }\href@noop {}
  {\bibfield  {journal} {\bibinfo  {journal} {Phys. Rev. Lett.}\ }\textbf
  {\bibinfo {volume} {111}},\ \bibinfo {pages} {070403} (\bibinfo {year}
  {2013})}\BibitemShut {NoStop}%
\bibitem [{\citenamefont {Yue}\ \emph {et~al.}(2014)\citenamefont {Yue},
  \citenamefont {Zhang},\ and\ \citenamefont {Fan}}]{Yue:2014uz}%
  \BibitemOpen
  \bibfield  {author} {\bibinfo {author} {\bibfnamefont {J.~D.}\ \bibnamefont
  {Yue}}, \bibinfo {author} {\bibfnamefont {Y.~R.}\ \bibnamefont {Zhang}}, \
  and\ \bibinfo {author} {\bibfnamefont {H.}~\bibnamefont {Fan}},\ }\href@noop
  {} {\bibfield  {journal} {\bibinfo  {journal} {Scientific Reports}\ }\textbf
  {\bibinfo {volume} {4}},\ \bibinfo {pages} {5933} (\bibinfo {year}
  {2014})}\BibitemShut {NoStop}%
\bibitem [{\citenamefont {Gagatsos}\ \emph {et~al.}(2016)\citenamefont
  {Gagatsos}, \citenamefont {Branford},\ and\ \citenamefont
  {Datta}}]{Gagatsos:2016ff}%
  \BibitemOpen
  \bibfield  {author} {\bibinfo {author} {\bibfnamefont {C.~N.}\ \bibnamefont
  {Gagatsos}}, \bibinfo {author} {\bibfnamefont {D.}~\bibnamefont {Branford}},
  \ and\ \bibinfo {author} {\bibfnamefont {A.}~\bibnamefont {Datta}},\
  }\href@noop {} {\bibfield  {journal} {\bibinfo  {journal} {Phys. Rev. A}\
  }\textbf {\bibinfo {volume} {94}},\ \bibinfo {pages} {042342} (\bibinfo
  {year} {2016})}\BibitemShut {NoStop}%
\bibitem [{\citenamefont {Knott}\ \emph {et~al.}(2016)\citenamefont {Knott},
  \citenamefont {Proctor}, \citenamefont {Hayes}, \citenamefont {Ralph},
  \citenamefont {Kok},\ and\ \citenamefont {Dunningham}}]{Knott:2016ig}%
  \BibitemOpen
  \bibfield  {author} {\bibinfo {author} {\bibfnamefont {P.~A.}\ \bibnamefont
  {Knott}}, \bibinfo {author} {\bibfnamefont {T.~J.}\ \bibnamefont {Proctor}},
  \bibinfo {author} {\bibfnamefont {A.~J.}\ \bibnamefont {Hayes}}, \bibinfo
  {author} {\bibfnamefont {J.~F.}\ \bibnamefont {Ralph}}, \bibinfo {author}
  {\bibfnamefont {P.}~\bibnamefont {Kok}}, \ and\ \bibinfo {author}
  {\bibfnamefont {J.~A.}\ \bibnamefont {Dunningham}},\ }\href@noop {}
  {\bibfield  {journal} {\bibinfo  {journal} {Phys. Rev. A}\ }\textbf {\bibinfo
  {volume} {94}},\ \bibinfo {pages} {062312} (\bibinfo {year}
  {2016})}\BibitemShut {NoStop}%
\bibitem [{\citenamefont {Ciampini}\ \emph {et~al.}(2016)\citenamefont
  {Ciampini}, \citenamefont {Spagnolo}, \citenamefont {Vitelli}, \citenamefont
  {Pezz{\`e}}, \citenamefont {Smerzi},\ and\ \citenamefont
  {Sciarrino}}]{Ciampini:2016cf}%
  \BibitemOpen
  \bibfield  {author} {\bibinfo {author} {\bibfnamefont {M.~A.}\ \bibnamefont
  {Ciampini}}, \bibinfo {author} {\bibfnamefont {N.}~\bibnamefont {Spagnolo}},
  \bibinfo {author} {\bibfnamefont {C.}~\bibnamefont {Vitelli}}, \bibinfo
  {author} {\bibfnamefont {L.}~\bibnamefont {Pezz{\`e}}}, \bibinfo {author}
  {\bibfnamefont {A.}~\bibnamefont {Smerzi}}, \ and\ \bibinfo {author}
  {\bibfnamefont {F.}~\bibnamefont {Sciarrino}},\ }\href@noop {} {\bibfield
  {journal} {\bibinfo  {journal} {Scientific Reports}\ }\textbf {\bibinfo
  {volume} {6}},\ \bibinfo {pages} {28881} (\bibinfo {year}
  {2016})}\BibitemShut {NoStop}%
\bibitem [{\citenamefont {Liu}\ \emph {et~al.}(2016)\citenamefont {Liu},
  \citenamefont {Lu}, \citenamefont {Sun},\ and\ \citenamefont
  {Wang}}]{Liu:2016by}%
  \BibitemOpen
  \bibfield  {author} {\bibinfo {author} {\bibfnamefont {J.}~\bibnamefont
  {Liu}}, \bibinfo {author} {\bibfnamefont {X.-M.}\ \bibnamefont {Lu}},
  \bibinfo {author} {\bibfnamefont {Z.}~\bibnamefont {Sun}}, \ and\ \bibinfo
  {author} {\bibfnamefont {X.}~\bibnamefont {Wang}},\ }\href@noop {} {\bibfield
   {journal} {\bibinfo  {journal} {Journal of Physics A: Mathematical and
  Theoretical}\ }\textbf {\bibinfo {volume} {49}},\ \bibinfo {pages} {115302}
  (\bibinfo {year} {2016})}\BibitemShut {NoStop}%
\bibitem [{\citenamefont {Zhang}\ and\ \citenamefont
  {Chan}(2017)}]{Zhang:2017bd}%
  \BibitemOpen
  \bibfield  {author} {\bibinfo {author} {\bibfnamefont {L.}~\bibnamefont
  {Zhang}}\ and\ \bibinfo {author} {\bibfnamefont {K.~W.~C.}\ \bibnamefont
  {Chan}},\ }\href@noop {} {\bibfield  {journal} {\bibinfo  {journal} {Phys.
  Rev. A}\ }\textbf {\bibinfo {volume} {95}},\ \bibinfo {pages} {032321}
  (\bibinfo {year} {2017})}\BibitemShut {NoStop}%
\bibitem [{\citenamefont {Ge}\ \emph {et~al.}(2017)\citenamefont {Ge},
  \citenamefont {Jacobs}, \citenamefont {Eldredge}, \citenamefont {Gorshkov},\
  and\ \citenamefont {Foss-Feig}}]{Ge:2017wf}%
  \BibitemOpen
  \bibfield  {author} {\bibinfo {author} {\bibfnamefont {W.}~\bibnamefont
  {Ge}}, \bibinfo {author} {\bibfnamefont {K.}~\bibnamefont {Jacobs}}, \bibinfo
  {author} {\bibfnamefont {Z.}~\bibnamefont {Eldredge}}, \bibinfo {author}
  {\bibfnamefont {A.~V.}\ \bibnamefont {Gorshkov}}, \ and\ \bibinfo {author}
  {\bibfnamefont {M.}~\bibnamefont {Foss-Feig}},\ }\href@noop {} {\  (\bibinfo
  {year} {2017})},\ \Eprint {http://arxiv.org/abs/1707.06655} {1707.06655}
  \BibitemShut {NoStop}%
\bibitem [{\citenamefont {Giovannetti}\ \emph {et~al.}(2011)\citenamefont
  {Giovannetti}, \citenamefont {Lloyd},\ and\ \citenamefont
  {Maccone}}]{Giovannetti:2011jk}%
  \BibitemOpen
  \bibfield  {author} {\bibinfo {author} {\bibfnamefont {V.}~\bibnamefont
  {Giovannetti}}, \bibinfo {author} {\bibfnamefont {S.}~\bibnamefont {Lloyd}},
  \ and\ \bibinfo {author} {\bibfnamefont {L.}~\bibnamefont {Maccone}},\
  }\href@noop {} {\bibfield  {journal} {\bibinfo  {journal} {Nature Photonics}\
  }\textbf {\bibinfo {volume} {5}},\ \bibinfo {pages} {222} (\bibinfo {year}
  {2011})}\BibitemShut {NoStop}%
\bibitem [{\citenamefont {Conde}(2017)}]{Conde:2017it}%
  \BibitemOpen
  \bibfield  {author} {\bibinfo {author} {\bibfnamefont {M.~H.}\ \bibnamefont
  {Conde}},\ }in\ \href@noop {} {\emph {\bibinfo {booktitle} {Compressive
  Sensing for the Photonic Mixer Device}}}\ (\bibinfo  {publisher} {Springer
  Vieweg, Wiesbaden},\ \bibinfo {address} {Wiesbaden},\ \bibinfo {year}
  {2017})\ pp.\ \bibinfo {pages} {11--88}\BibitemShut {NoStop}%
\bibitem [{\citenamefont {Lodewyck}\ \emph {et~al.}(2009)\citenamefont
  {Lodewyck}, \citenamefont {Westergaard},\ and\ \citenamefont
  {Lemonde}}]{Lodewyck:2009ib}%
  \BibitemOpen
  \bibfield  {author} {\bibinfo {author} {\bibfnamefont {J.}~\bibnamefont
  {Lodewyck}}, \bibinfo {author} {\bibfnamefont {P.~G.}\ \bibnamefont
  {Westergaard}}, \ and\ \bibinfo {author} {\bibfnamefont {P.}~\bibnamefont
  {Lemonde}},\ }\href@noop {} {\bibfield  {journal} {\bibinfo  {journal} {Phys.
  Rev. A}\ }\textbf {\bibinfo {volume} {79}},\ \bibinfo {pages} {061401}
  (\bibinfo {year} {2009})}\BibitemShut {NoStop}%
\bibitem [{\citenamefont {Weedbrook}\ \emph {et~al.}(2012)\citenamefont
  {Weedbrook}, \citenamefont {Pirandola}, \citenamefont
  {Garc{\'\i}a-Patr{\'o}n}, \citenamefont {Cerf}, \citenamefont {Ralph},
  \citenamefont {Shapiro},\ and\ \citenamefont {Lloyd}}]{Weedbrook:2012tz}%
  \BibitemOpen
  \bibfield  {author} {\bibinfo {author} {\bibfnamefont {C.}~\bibnamefont
  {Weedbrook}}, \bibinfo {author} {\bibfnamefont {S.}~\bibnamefont
  {Pirandola}}, \bibinfo {author} {\bibfnamefont {R.}~\bibnamefont
  {Garc{\'\i}a-Patr{\'o}n}}, \bibinfo {author} {\bibfnamefont {N.~J.}\
  \bibnamefont {Cerf}}, \bibinfo {author} {\bibfnamefont {T.~C.}\ \bibnamefont
  {Ralph}}, \bibinfo {author} {\bibfnamefont {J.~H.}\ \bibnamefont {Shapiro}},
  \ and\ \bibinfo {author} {\bibfnamefont {S.}~\bibnamefont {Lloyd}},\
  }\href@noop {} {\bibfield  {journal} {\bibinfo  {journal} {Reviews of Modern
  Physics}\ }\textbf {\bibinfo {volume} {84}},\ \bibinfo {pages} {621}
  (\bibinfo {year} {2012})}\BibitemShut {NoStop}%
\bibitem [{\citenamefont {Holevo}(1982)}]{holevoprob}%
  \BibitemOpen
  \bibfield  {author} {\bibinfo {author} {\bibfnamefont {A.~S.}\ \bibnamefont
  {Holevo}},\ }\href@noop {} {\emph {\bibinfo {title} {Probabilistic and
  Statistical Aspects of Quantum Theory}}}\ (\bibinfo  {publisher}
  {North-Holland, Amsterdam},\ \bibinfo {year} {1982})\BibitemShut {NoStop}%
\bibitem [{\citenamefont {Jarzyna}\ and\ \citenamefont
  {Demkowicz-Dobrza\ifmmode~\acute{n}\else
  \'{n}\fi{}ski}(2012)}]{jarzynawithandwithout}%
  \BibitemOpen
  \bibfield  {author} {\bibinfo {author} {\bibfnamefont {M.}~\bibnamefont
  {Jarzyna}}\ and\ \bibinfo {author} {\bibfnamefont {R.}~\bibnamefont
  {Demkowicz-Dobrza\ifmmode~\acute{n}\else \'{n}\fi{}ski}},\ }\href@noop {}
  {\bibfield  {journal} {\bibinfo  {journal} {Phys. Rev. A}\ }\textbf {\bibinfo
  {volume} {85}},\ \bibinfo {pages} {011801} (\bibinfo {year}
  {2012})}\BibitemShut {NoStop}%
\bibitem [{Note1()}]{Note1}%
  \BibitemOpen
  \bibinfo {note} {As Ref. \cite {jarzynawithandwithout} points out, the SQL
  for phase difference estimation depends on whether we assume access to a
  phase reference or not. In our setting, the receiver has a local oscillator
  that provides a phase reference (\protect \emph {e.g.,}~has a fixed phase
  relationship to $\rho _{\protect \rm in}$) and hence the SQL in this setting
  is $\alpha _1^2 + \alpha _2^2$.}\BibitemShut {Stop}%
\bibitem [{\citenamefont {Nichols}\ \emph {et~al.}(2018)\citenamefont
  {Nichols}, \citenamefont {Liuzzo-Scorpo}, \citenamefont {Knott},\ and\
  \citenamefont {Adesso}}]{adessomulti}%
  \BibitemOpen
  \bibfield  {author} {\bibinfo {author} {\bibfnamefont {R.}~\bibnamefont
  {Nichols}}, \bibinfo {author} {\bibfnamefont {P.}~\bibnamefont
  {Liuzzo-Scorpo}}, \bibinfo {author} {\bibfnamefont {P.~A.}\ \bibnamefont
  {Knott}}, \ and\ \bibinfo {author} {\bibfnamefont {G.}~\bibnamefont
  {Adesso}},\ }\href {\doibase 10.1103/PhysRevA.98.012114} {\bibfield
  {journal} {\bibinfo  {journal} {Phys. Rev. A}\ }\textbf {\bibinfo {volume}
  {98}},\ \bibinfo {pages} {012114} (\bibinfo {year} {2018})}\BibitemShut
  {NoStop}%
\bibitem [{\citenamefont {Serafini}(2017)}]{serafinibook}%
  \BibitemOpen
  \bibfield  {author} {\bibinfo {author} {\bibfnamefont {A.}~\bibnamefont
  {Serafini}},\ }\href@noop {} {\emph {\bibinfo {title} {Quantum continuous
  variables: a primer of theoretical methods}}}\ (\bibinfo  {publisher} {CRC
  {Press}, {Taylor \& Francis} {Group}},\ \bibinfo {year} {2017})\BibitemShut
  {NoStop}%
\bibitem [{\citenamefont {Braunstein}\ and\ \citenamefont
  {Caves}(1994)}]{braunsteincaves}%
  \BibitemOpen
  \bibfield  {author} {\bibinfo {author} {\bibfnamefont {S.~L.}\ \bibnamefont
  {Braunstein}}\ and\ \bibinfo {author} {\bibfnamefont {C.~M.}\ \bibnamefont
  {Caves}},\ }\href@noop {} {\bibfield  {journal} {\bibinfo  {journal} {Phys.
  Rev. Lett.}\ }\textbf {\bibinfo {volume} {72}},\ \bibinfo {pages} {3439}
  (\bibinfo {year} {1994})}\BibitemShut {NoStop}%
\bibitem [{\citenamefont {Wiseman}(1996)}]{Wis-1996}%
  \BibitemOpen
  \bibfield  {author} {\bibinfo {author} {\bibfnamefont {H.~M.}\ \bibnamefont
  {Wiseman}},\ }\href@noop {} {\bibfield  {journal} {\bibinfo  {journal}
  {Quantum Semiclass. Opt.}\ }\textbf {\bibinfo {volume} {8}},\ \bibinfo
  {pages} {205} (\bibinfo {year} {1996})}\BibitemShut {NoStop}%
\bibitem [{\citenamefont {Monras}(2006)}]{monrasphase}%
  \BibitemOpen
  \bibfield  {author} {\bibinfo {author} {\bibfnamefont {A.}~\bibnamefont
  {Monras}},\ }\href@noop {} {\bibfield  {journal} {\bibinfo  {journal} {Phys.
  Rev. A}\ }\textbf {\bibinfo {volume} {73}},\ \bibinfo {pages} {033821}
  (\bibinfo {year} {2006})}\BibitemShut {NoStop}%
\bibitem [{\citenamefont {\ifmmode~\check{S}\else \v{S}\fi{}afr\'anek}\ and\
  \citenamefont {Fuentes}(2016)}]{safranekfuentes}%
  \BibitemOpen
  \bibfield  {author} {\bibinfo {author} {\bibfnamefont {D.}~\bibnamefont
  {\ifmmode~\check{S}\else \v{S}\fi{}afr\'anek}}\ and\ \bibinfo {author}
  {\bibfnamefont {I.}~\bibnamefont {Fuentes}},\ }\href {\doibase
  10.1103/PhysRevA.94.062313} {\bibfield  {journal} {\bibinfo  {journal} {Phys.
  Rev. A}\ }\textbf {\bibinfo {volume} {94}},\ \bibinfo {pages} {062313}
  (\bibinfo {year} {2016})}\BibitemShut {NoStop}%
\bibitem [{\citenamefont {Dutta}\ \emph {et~al.}(1995)\citenamefont {Dutta},
  \citenamefont {Mukunda},\ and\ \citenamefont {Simon}}]{Dut.Muk.etal-1995}%
  \BibitemOpen
  \bibfield  {author} {\bibinfo {author} {\bibfnamefont {A.~B.}\ \bibnamefont
  {Dutta}}, \bibinfo {author} {\bibfnamefont {N.}~\bibnamefont {Mukunda}}, \
  and\ \bibinfo {author} {\bibfnamefont {R.}~\bibnamefont {Simon}},\
  }\href@noop {} {\bibfield  {journal} {\bibinfo  {journal} {Pramana}\ }\textbf
  {\bibinfo {volume} {45}},\ \bibinfo {pages} {471} (\bibinfo {year}
  {1995})}\BibitemShut {NoStop}%
\bibitem [{Note2()}]{Note2}%
  \BibitemOpen
  \bibinfo {note} {While the GFI does not depend on the values of the
  parameters, the form of the optimal measurement does.}\BibitemShut {Stop}%
\bibitem [{\citenamefont {Caruso}\ \emph {et~al.}(2006)\citenamefont {Caruso},
  \citenamefont {Giovanetti},\ and\ \citenamefont {Holevo}}]{holevogiov}%
  \BibitemOpen
  \bibfield  {author} {\bibinfo {author} {\bibfnamefont {F.}~\bibnamefont
  {Caruso}}, \bibinfo {author} {\bibfnamefont {V.}~\bibnamefont {Giovanetti}},
  \ and\ \bibinfo {author} {\bibfnamefont {A.~S.}\ \bibnamefont {Holevo}},\
  }\href@noop {} {\bibfield  {journal} {\bibinfo  {journal} {New J. Phys.}\
  }\textbf {\bibinfo {volume} {8}},\ \bibinfo {pages} {310} (\bibinfo {year}
  {2006})}\BibitemShut {NoStop}%
\bibitem [{\citenamefont
  {Demkowicz-Dobrza{\'{n}}ski}(2011)}]{DemkowiczDobrzanski:2011ko}%
  \BibitemOpen
  \bibfield  {author} {\bibinfo {author} {\bibfnamefont {R.}~\bibnamefont
  {Demkowicz-Dobrza{\'{n}}ski}},\ }\href@noop {} {\bibfield  {journal}
  {\bibinfo  {journal} {Phys. Rev. A}\ }\textbf {\bibinfo {volume} {83}},\
  \bibinfo {pages} {061802} (\bibinfo {year} {2011})}\BibitemShut {NoStop}%
\bibitem [{Note3()}]{Note3}%
  \BibitemOpen
  \bibinfo {note} {It should be noted that although the measurement is
  independent of the actual phase values, the form of the estimation algorithm
  (\protect \emph {e.g.,}~maximal likelihood estimator) that saturates the
  Cramer-Rao bound may still depend on the phase values.}\BibitemShut {Stop}%
\end{thebibliography}%

\end{document}